\documentclass[10pt,journal,compsoc]{IEEEtran}
\ifCLASSOPTIONcompsoc
  \usepackage[nocompress]{cite}
\else
  \usepackage{cite}
\fi
\ifCLASSINFOpdf
  \usepackage[pdftex]{graphicx}
\else
  \usepackage[dvips]{graphicx}
\fi
\usepackage{amsmath}
\usepackage{algorithmic}
\usepackage{array}
\usepackage{url}
 
\usepackage[american]{babel}
\usepackage{microtype}
\usepackage{multirow}
\usepackage{multicol}
\usepackage{tabularx}
\usepackage{array}
\usepackage{makecell}
\usepackage{flushend}
\usepackage[linesnumbered,ruled]{algorithm2e}
\usepackage[draft]{hyperref}
\usepackage{braket}
\usepackage[usenames, dvipsnames]{color}
\usepackage{pifont}
\usepackage{ulem}

\begin{document}
\normalem
\title{\fontsize{23}{25}\selectfont QuCloud+: A Holistic Qubit Mapping Scheme for Single/Multi-programming on 2D/3D NISQ Quantum Computers}
%
%
%
%

\author{Lei~Liu, Xinglei~Dou
\IEEEcompsocitemizethanks{
\IEEEcompsocthanksitem Lei Liu (corresponding author), Xinglei Dou are with the Sys-Inventor Lab, Beihang University; SKLCA, ICT, CAS; No.6 Kexueyuan South Road Zhongguancun, Haidian District Beijing, China.\protect\\
E-mail: lei.liu@zoho.com; liulei2010@{buaa.edu.cn/ict.ac.cn}
}
\thanks{}}

%
%

\markboth{}%
{Liu \MakeLowercase{\textit{et al.}}: QuCloud: A New Qubit Mapping Mechanism for Multi-programming Quantum Computing}
%




\IEEEtitleabstractindextext{%
\begin{abstract}
Qubit mapping is essential to quantum computing's fidelity and quantum computers' resource utilization. Yet, the existing qubit mapping schemes meet some challenges (e.g., crosstalk, SWAP overheads, diverse device topologies, etc.), leading to qubit resource under-utilization, high error rate, and low fidelity in computing results. This paper presents QuCloud+, a new qubit mapping scheme capable of handling these challenges. QuCloud+ has several new designs. (1) QuCloud+ enables multi-programming quantum computing on quantum chips with 2D/3D topology. (2) It partitions physical qubits for concurrent quantum programs with the crosstalk-aware community detection technique and further allocates qubits according to qubit degree, improving fidelity and resource utilization. (3) QuCloud+ includes an X-SWAP mechanism that avoids SWAPs with high crosstalk errors and enables inter-program SWAPs to reduce the SWAP overheads. (4) QuCloud+ schedules concurrent quantum programs to be mapped and executed based on estimated fidelity for the best practice. QuCloud+ outperforms the previous multi-programming work on various devices by 6.84\% on fidelity and saves 40.9\% additional gates required during mapping transition. 
\end{abstract}

}

\maketitle

\IEEEdisplaynontitleabstractindextext

%
\IEEEpeerreviewmaketitle

\IEEEraisesectionheading{\section{Introduction}\label{sec:introduction}}
\IEEEPARstart{Q}{uantum} computers have gradually entered our field of vision. Due to its potential in various critical applications, such as machine learning~\cite{c25_ml,c26_ml}, database search~\cite{c3_dbsearch} and chemistry simulation~\cite{c4_simulator,c24_chem}, many companies, universities and institutes are actively working to develop prototypes of quantum computer systems. Recently, quantum devices with tens of quantum bits (qubits) are delivered by Google~\cite{c7_google72q}, Intel~\cite{c5_intel49q}, and IBM~\cite{c6_ibm53q,c27_ibmq50}, etc. Modern quantum computers belong to the Noisy Intermediate-Scale Quantum (NISQ) category~\cite{c1_beyond} – the qubits and the links between them are with variational reliability and are easily disturbed; therefore, quantum computers are susceptible to errors. Several competing qubit technologies are available for the physical implementation of quantum devices, e.g., trapped ion qubits~\cite{c17_ion}, superconducting qubits~\cite{c16_superconducting}, silicon qubits~\cite{c32_silicon}, and photonic qubits~\cite{jiuzhang}. Among these technologies, using superconducting qubit is promising \cite{c15_ibmq,c8_multi,c12_SABRE,c52_qucloud,c71_PACT}. This work mainly focuses on IBM QX architectures with superconducting qubits.

Quantum circuit (program) mapping is a fundamental mechanism in NISQ era. The quantum computing software system (e.g., OS, compiler) \cite{c8_multi,c12_SABRE,c52_qucloud} makes the quantum circuits compatible with the target quantum chips and maps them onto the chips. Some qubit mapping mechanisms typically map only a single quantum program for higher fidelity~\cite{c12_SABRE,c9_noise_adaptive,c11_not_all}. Recent studies~\cite{c8_multi,c52_qucloud,c57_VQA,c71_PACT} introduce multi-programming in NISQ computers for higher throughput and resource utilization of quantum computers. Enabling multi-programming can effectively improve the utilization of robust qubits on quantum computers, improve the throughput of quantum computing cloud services, and speed up variational quantum algorithms (VQAs)~\cite{c57_VQA}. Although multi-programming on quantum computers may have many benefits, it also brings new challenges. One notable problem is that the activity of a program can negatively affect the reliability of co-located programs because of (\romannumeral1) a limited number of robust qubits, (\romannumeral2) crosstalk noise caused by unexpected interactions or in-corrected control of qubits~\cite{c21_crosstalk,c60_crosstalk} and (\romannumeral3) long qubit SWAP paths. Previous study~\cite{c8_multi} on multi-programming shows that running 2 quantum programs on a specific quantum chip incurs a 12.0\% reduction on fidelity, on average.

Qubit mapping can be a key component in quantum computing stack. We find the previous qubit mapping policies have several shortcomings when handling multi-programming cases. (1) The existing mapping policies often divide a large area of robust on-chip qubits into many small-scale segments that other programs cannot map onto. In many cases, over 20\% of the robust qubits are wasted during the initial mapping. (2) When a specific quantum chip is partitioned for mapping multiple quantum programs, SWAP operations required during mapping transition for each quantum program increase, leading to an unpredictable impact for fidelity and reliability. (3) Crosstalk errors \cite{c60_crosstalk,c21_crosstalk} often occur when executing concurrent quantum programs. (4) Scheduling concurrent quantum 
programs on a specific quantum chip is a challenging job, leading to fidelity degradation and resource under-utilization in the cloud environment ~\cite{c52_qucloud,c8_multi}.

To this end, in this paper, we propose a new approach, i.e., QuCloud+ based on prior studies \cite{c52_qucloud,c71_PACT}, to improve the throughput and resource utilization of NISQ machines in the cloud environment while reducing the negative impacts of multi-programming on NISQ computers' reliability. QuCloud+ has several key features. (1) QuCloud+ supports both single-program mapping and multi-program mapping, and can switch across the two modes on demand. (2) QuCloud+ is crosstalk-aware. It mitigates crosstalk errors by orchestrating the qubit allocations and SWAP operations (Sec. \ref{part:CDAP}, \ref{part:xswap}). Using these technologies, QuCloud+ improves the fidelity in latest studies on QuCloud \cite{c52_qucloud,c71_PACT} by 0.49\%. (3) QuCloud+ partitions the physical qubits for concurrent quantum programs leveraging community detection technique~\cite{c20_FN}, avoiding the waste caused by the topology-unaware algorithms. (4) QuCloud+ can enable inter-program SWAPs when concurrent quantum programs are allocated to neighboring qubits, which reduces the overall SWAP overheads. (5) QuCloud+ supports both 2D and 3D quantum chips. Experimental results show that QuCloud+ outperforms the latest multi-programming solution~\cite{c8_multi} by 6.84\% on fidelity and saves 40.9\% additional gates required for multi-programming workloads.

To sum up, in QuCloud+, we make the following contributions. (1) We design a new qubit mapping scheme, including a new approach (CDAP) that allocates robust qubit sets for concurrent quantum programs and a mapping transition mechanism (X-SWAP) that enables inter-program SWAP when quantum programs are mapped adjacently, significantly reducing SWAP overheads. (2) We further enhance our design based on newly proposed quantum program profiling approach, and show QuCloud+ has advantages on 2D quantum chip as well as the emerging 3D quantum chips. (3) Regarding crosstalk mitigation, we design crosstalk-aware initial mapping generation and crosstalk-aware mapping transition, mitigating the crosstalk problem on quantum devices obviously. (4) We design the QuCloud+ scheduler that selects appropriate quantum program combinations for multi-programming. Moreover, using the new scheduler, QuCloud+ can switch between single-program mapping and multi-program mapping easily on demand.

\begin{figure}[t]
\centering
\includegraphics[width=0.9\linewidth]{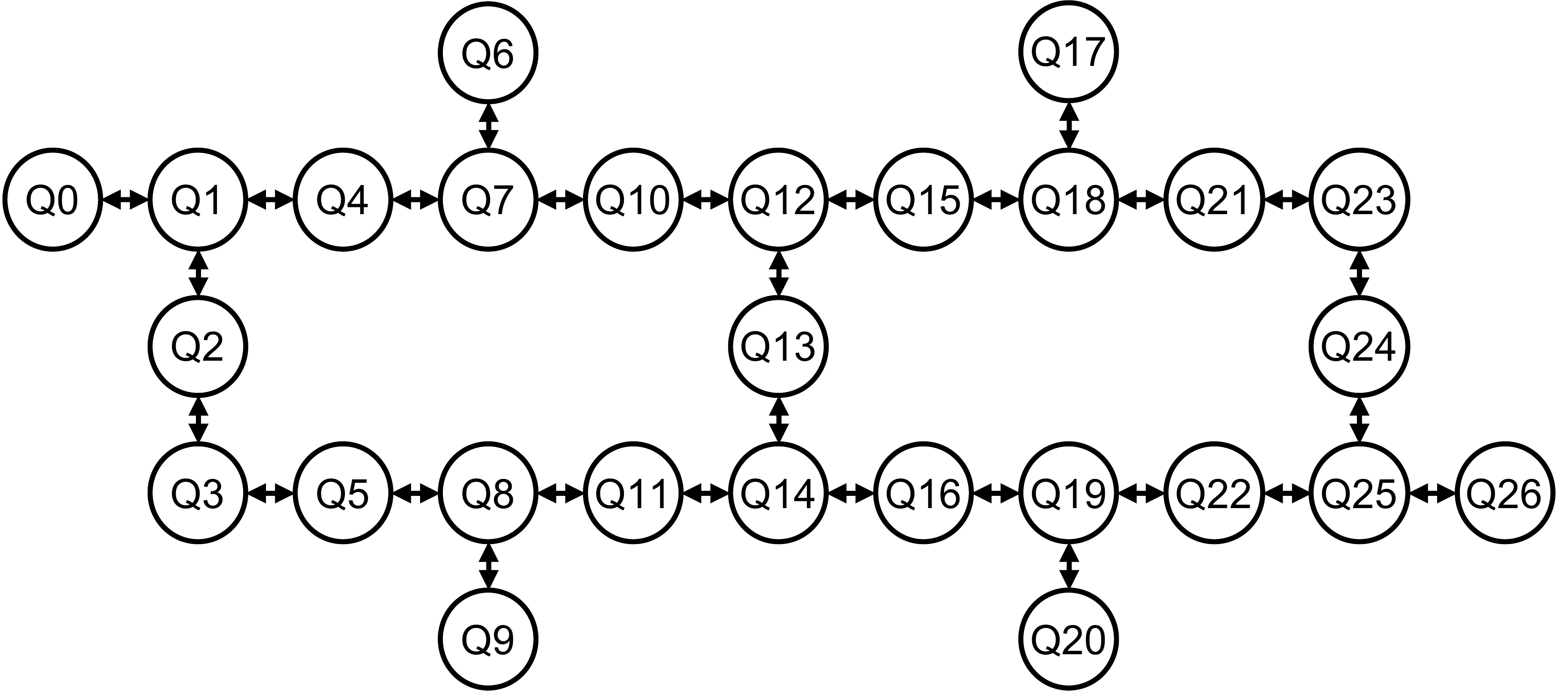}
\vspace{-0.2cm}
\caption{Architecture of IBMQ toronto.}
\label{fig:2}
\end{figure}

\section{Background}

\subsection{Quantum Computing Basics}
Quantum computing can solve conventionally hard problems leveraging quantum mechanism~\cite{c2_factorization}. The foundation of quantum computing lies on qubits. $\ket{0}$ and $\ket{1}$ are two basis states of a qubit. The state of a qubit can be the linear combinations of $\ket{0}$ and $\ket{1}$ as $|\psi\rangle = \alpha \ket{0} + \beta \ket{1}$, where $\alpha$, $\beta$ are complex numbers with $|\alpha|^2 + |\beta|^2 = 1$. The state of a qubit can be manipulated with single-qubit gates, e.g., \emph{H}, \emph{X}, \emph{Y}, \emph{Z}, etc. When the qubit is measured, the measurement result gives either 0 with the probability of $|\alpha|^2$, or 1 with the probability of $|\beta|^2$. Two or more qubits can be entangled with two-qubit gates, i.e., Control-NOT (CNOT) gates. A CNOT gate flips the state of the target qubit when the control qubit is in the state $\ket{1}$. Likewise, the state of a two-qubit system is represented by: $|\psi\rangle = \alpha_{00} \ket{00} + \alpha_{01} \ket{01} +\alpha_{10} \ket{10} +\alpha_{11} \ket{11}$. Any quantum gate involving three or more qubits can be decomposed with single- and two-qubit gates~\cite{c19_decompose}.

\subsection{Quantum Computers} IBM quantum processors are superconducting quantum chips with Josephson-junction-based transmon qubits~\cite{c16_superconducting} and microwave-tunable two-qubit gates~\cite{c18_microwave}. Unlike the all-to-all connectivity of qubits on ion-trap quantum computers, the physical qubits on IBM quantum chips only have connections to neighboring qubits. Figure~\ref{fig:2} illustrates the architecture of IBMQ toronto.

\subsection{Errors on Quantum Computers} 
NISQ computers have to face reliability challenges. As the physical qubits are fragile and susceptible to interference, the following kinds of errors may occur in quantum programs running on quantum computers. (\romannumeral1) Coherence errors caused by short qubit state retention time~\cite{c15_ibmq}. (\romannumeral2) Operational errors caused by error-prone quantum gates~\cite{c15_ibmq}. (\romannumeral3) Readout errors caused by measurement operations~\cite{c15_ibmq}. (\romannumeral4) Crosstalk errors caused by unexpected interactions between quantum gates or imprecise quantum control when quantum gates are executed simultaneously~\cite{c69_crosstalk}. In practice, operational error rate, readout error rate, and the coherence time of an IBM backend are all reported in its calibration data~\cite{c15_ibmq}. The error rates vary for different qubits, links, and days. Crosstalk errors can be measured using Simultaneous Randomized Benchmarking (SRB)~\cite{c70_SRB}. In practice, quantum programs should be mapped to qubits with lower error rates.

\begin{figure}[t]
\centering
\includegraphics[width=0.80\linewidth]{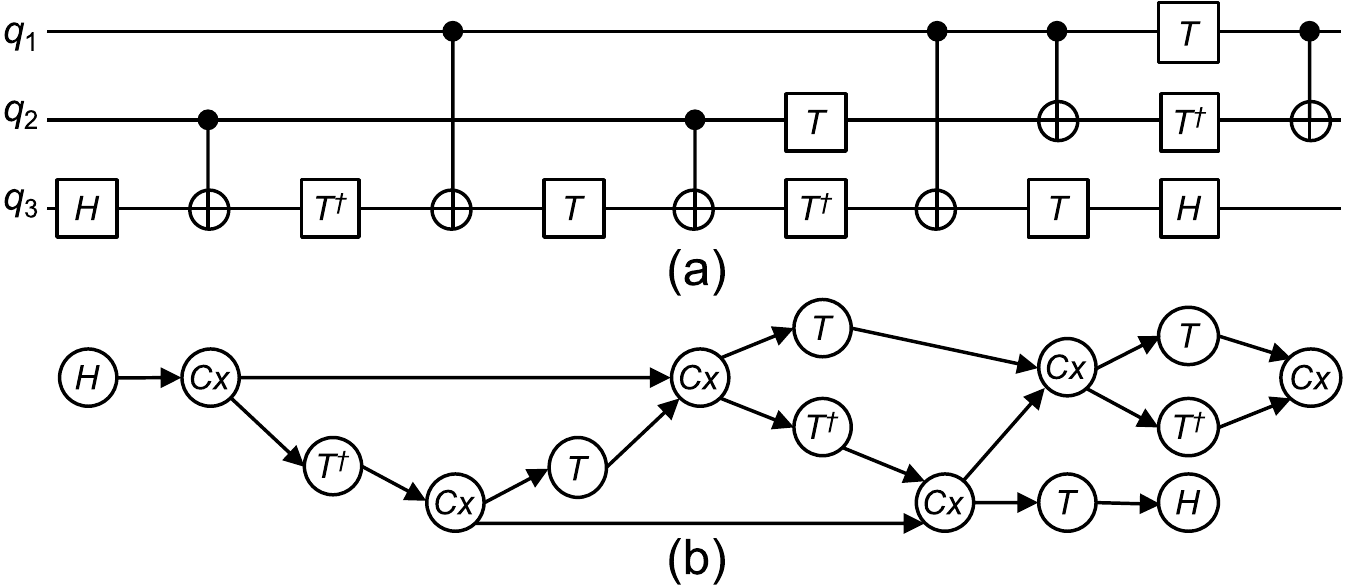}
\vspace{-0.2cm}
\caption{The Quantum circuit and DAG of decomposed Toffoli gate.}
\label{fig:3}
\end{figure}

\subsection{Quantum Programs and Quantum Circuits}
Quantum programs can be converted into quantum circuits composed of a series of quantum gates. For example, Figure~\ref{fig:3}-(a) shows the decomposed Toffoli gate's quantum circuit, where each horizontal line represents a program qubit, each block and vertical line represents a single- and two-qubit gate, respectively. Figure~\ref{fig:3}-(b) shows the Directed Acyclic Graph (DAG) of the circuit. The depth of the circuit equals the length of the DAG's critical path. A quantum gate is logically executable when it has no unexecuted predecessors in the DAG. A CNOT operation cannot be executed unless the two program qubits involved are mapped physically adjacent. A SWAP operation exchanges the physical mapping between two program qubits \cite{c12_SABRE,c31_intro}. Generally, it takes two steps for a heuristic-based policy to solve the mapping problem. (1) Initial mapping generation. The policy maps each program's program qubits onto physical qubits. (2) Mapping transition. The policy meets all two-qubit constraints by inserting SWAPs to the quantum circuit so that every two-qubit gate in the quantum program can be executed physically.

\section{Motivations}

\subsection{New Architecture, Paradigm and System Stack}
\subsubsection{Quantum Chip with 3D Topology}
Superconducting quantum chips are often with 2D topology architectures. Their physical qubits are organized in a lattice layout on a 2D plane \cite{c67_design,c52_qucloud,c71_PACT}. Connections only exist between neighboring physical qubits. SWAPs are required when two program qubits involved in a quantum gate are not mapped on neighboring physical qubits with interconnection. Some studies~\cite{c61_dense,c62_25d,c64_3d,c65_3d} show that quantum computing benefits from the quantum chip topology with a higher dimension, e.g., a 3D quantum chip whose qubits are organized in a 3D lattice layout. A quantum chip with a higher dimension has a denser qubit topological structure, higher physical qubit degrees, and more direct interconnections between its physical qubits. These features help reduce the additional SWAPs. Moreover, new advances in quantum chip manufacturing, e.g., qubit memory technology~\cite{c63_random}, qubit integration and packaging technology~\cite{c66_integration}, and quantum chip architecture design~\cite{c67_design}, also provide possibilities for the devices with a denser structure and a higher dimension in the near future. The new trends motivate us that we should have new designs on 3D quantum chips and have new mapping/allocation schemes for the new architectures.

\subsubsection{Multi-programming on Quantum Computers}
Multi-programming is introduced in~\cite{c8_multi} to improve the qubit utilization and the overall throughput for the expensive quantum computers. It takes full advantage of the limited robust resources of a specific quantum computer and can speed up certain quantum algorithms such as VQAs \cite{c57_VQA}. For the quantum computing service providers, multi-programming becomes more useful than ever as the number of qubits increases, and researchers worldwide want to use quantum computers. However, multi-programming on quantum computers is not a free lunch. Several studies \cite{c8_multi,c52_qucloud,c57_VQA,c71_PACT} provide solutions to optimize throughput and fidelity in multi-programming scenarios. In this paper, we devise new approaches to address the qubit mapping and crosstalk among concurrent quantum programs.

\subsubsection{A New OS for Quantum Computers}
It is the right time for the quantum computer to have a new OS – \emph{QuOS}. QuOS is the manager of both hardware and software resources in the quantum computer. It is responsible for managing and configuring the qubits with varying hardware implementations, scheduling quantum programs, and providing the best mapping strategies for higher fidelity. Our work is among the first step studies~\cite{c50_QOS_1,c51_QOS_2,c52_qucloud} that discuss the prototype of OS for quantum computers. The quantum computer has different design principles from the classical Von Neumann architecture. The critical OS components, e.g., ISA, scheduling, process management, etc., are incompatible with the quantum architecture. Therefore, OS's design principles and implementation strategies for quantum computers should be different from traditional ones. We are exploring OS and run-time stack technologies for quantum computing.

\subsection{Key Challenges -- What Should We Do?}
\subsubsection{Noise on Quantum Computers - Fidelity}
Quantum computers in the NISQ era are susceptible to errors. The state of a specific qubit can only keep in a short time (e.g., 30-100 $\mu$s)~\cite{c15_ibmq}. In reality, the error rates are variational across all of the qubits and links on NISQ computers. To map a specific quantum program, previous noise-aware mapping techniques for a single quantum program~\cite{c9_noise_adaptive,c10_extract,c11_not_all} employ greedy or heuristic approaches to discover the mapping policies that have the most reliable qubits and links.

Many prior studies \cite{c11_not_all,c9_noise_adaptive,c14_BKA,c52_qucloud} assume that errors on quantum computers are local and independent. However, crosstalk errors violate either the locality or the independence of quantum operations (or both) \cite{c69_crosstalk}. Crosstalk commonly exists on NISQ systems. Previous studies demonstrate that crosstalk errors mainly occur among CNOT pairs \cite{c21_crosstalk}. When two CNOT gates are executed simultaneously, the error rate of both CNOT gates is amplified. In this work, we use the error amplification ratio to indicate the effect of crosstalk, defined as $r_{i|j}=e_{i|j}/e_i$. $e_i$ denotes the error rate of gate $i$ when it is executed independently, $e_{i|j}$ denotes the error rate of gate $i$ when it is executed simultaneously with gate $j$.

\subsubsection{Qubit Mapping can be More Effective and Efficient}
Enabling multi-programming on a specific quantum chip brings challenges for existing qubit mapping technologies. (1) The available qubits are subject to be divided into smaller scale segments. Some of them are suitable to be involved during mapping, but some are not due to their high error rate and weak links. Hence, some programs might have weak qubits when mapping multiple quantum programs, leading to unreliable results and increasing error rate for concurrent quantum programs. (2) Although it is possible to combine multiple quantum programs into one quantum circuit and then map it using the mapping policies dedicated to single quantum programs, the following problems may occur. (i) Reliable resources (robust qubits) on a specific quantum chip are limited, and no fairness for allocating reliable qubits among co-located quantum programs is guaranteed. (ii) The number of concurrent quantum programs cannot be adjusted on-the-fly. For example, when a significant fidelity reduction happens for multi-programming, the parallel mode cannot be reverted to separate execution mode. (iii) Some optimization opportunities for multi-programming are missed. For example, the SWAP overheads can be reduced by leveraging inter-program SWAPs. If a program occupies any qubits on the shortest SWAP path for any other co-located program, the SWAP process has to suffer higher overheads, i.e., involving more SWAP operations across more qubits. More SWAP operations may bring higher error rate; Therefore, the overall fidelity is negatively affected.

The previous multi-programming technique~\cite{c8_multi} supports co-locating two quantum programs. Some studies propose solutions for mapping single quantum program~\cite{c11_not_all,c12_SABRE}. They mainly rely on heuristic policies or greedy algorithms. Yet, mapping multiple programs differs from the scheme that maps a single program; This inspires us to devise a new qubit mapping mechanism.

\subsection{Related Work}
The qubit mapping problem has been proved to be NP-Complete \cite{c53_QA,c54_OPT}. Recent studies on quantum program mapping are within two categories. (1) Optimal solutions \cite{c53_QA,c21_crosstalk,c54_OPT,c_optimal_1,c_optimal_2,c_optimal_3,c_optimal_4,c_optimal_5}. These approaches convert the qubit mapping problem to an equivalent mathematical optimization problem, then apply a solver to solve it. They can have the optimal solution for small-scale quantum program mapping problem but might suffer from a high time complexity. (2) Heuristic solutions \cite{c9_noise_adaptive,c14_BKA,c11_not_all,c12_SABRE,C30_QURE,c_heuristic_1,c_heuristic_2,c_heuristic_3}. Though heuristic approaches cannot ensure that the solution obtained is the optimal, they are more flexible. For example, the IBM Qiskit framework~\cite{c13_qiskit} implements noisy adaptive heuristic mapping~\cite{c9_noise_adaptive} and Stochastic SWAP. SABRE~\cite{c12_SABRE} brings exponential speedup in the search complexity by reducing the search space. However, most prior work focus on mapping problems for only one quantum program. New policies are required to map multiple quantum programs onto a specific quantum chip simultaneously. For multi-programming, the effort in~\cite{c8_multi} proposes the FRP algorithm to assign reliable regions for each quantum program. It enhances SABRE~\cite{c12_SABRE} with noise awareness to generate the mapping transition. In this paper, we devise a new mapping scheme for multi-programming with respect to mapping on 3D chip architecture, alleviating crosstalk errors, etc.

\begin{figure}[t]
\centering
\includegraphics[width=0.99\linewidth]{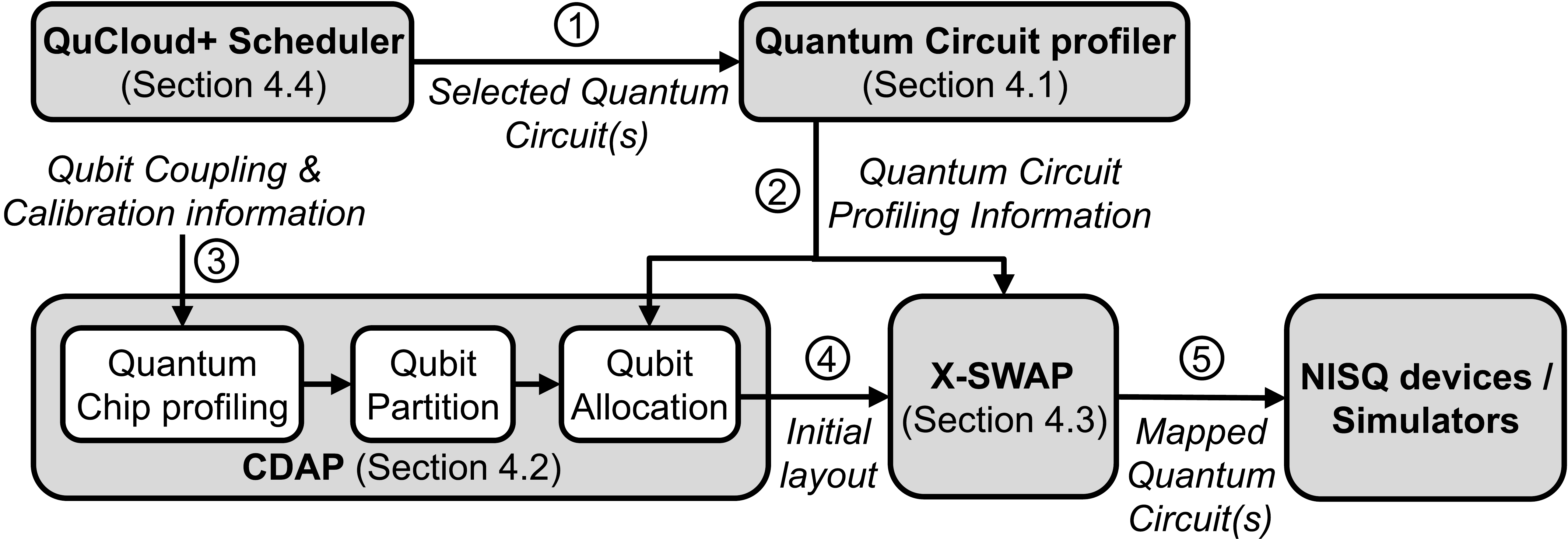}
\vspace{-0.2cm}
\caption{Overview of the QuCloud+ design.}
\label{fig:overview}
\end{figure}

\section{The Art of Our Design}
\textbf{Overview}. In this section, we show QuCloud+, which is a new qubit mapping scheme for multiple (or single) quantum programs in cloud. Figure \ref{fig:overview} illustrates the overview of our design. The QuCloud+ scheduler selects quantum circuits to be mapped in {\fontsize{12}{10}\selectfont\ding{172}}. Then, QuCloud+ identifies the CNOT patterns for these quantum circuits. It finds when each qubit is involved in consecutive CNOTs, and how long the qubit is involved. We denote this process as profiling in {\fontsize{12}{10}\selectfont\ding{173}}. The profiling results are used in the quantum circuit mapping process. Then, QuCloud+ generates the initial mapping using CDAP. CDAP clusters physical qubits according to the qubit coupling and calibration information in {\fontsize{12}{10}\selectfont\ding{174}}, providing guidance for qubit partitioning of concurrent quantum circuits. After each quantum circuit is virtually assigned to a set of physical qubits, QuCloud+ maps program qubits to physical qubits based on the quantum program profiling details from {\fontsize{12}{10}\selectfont\ding{173}}. Mapping transition is conducted in {\fontsize{12}{10}\selectfont\ding{175}} by enabling X-SWAP, and the mapped circuits are forward to NISQ devices to execute in {\fontsize{12}{10}\selectfont\ding{176}}. 

\begin{figure}[t]
\centering
\includegraphics[width=0.2\linewidth]{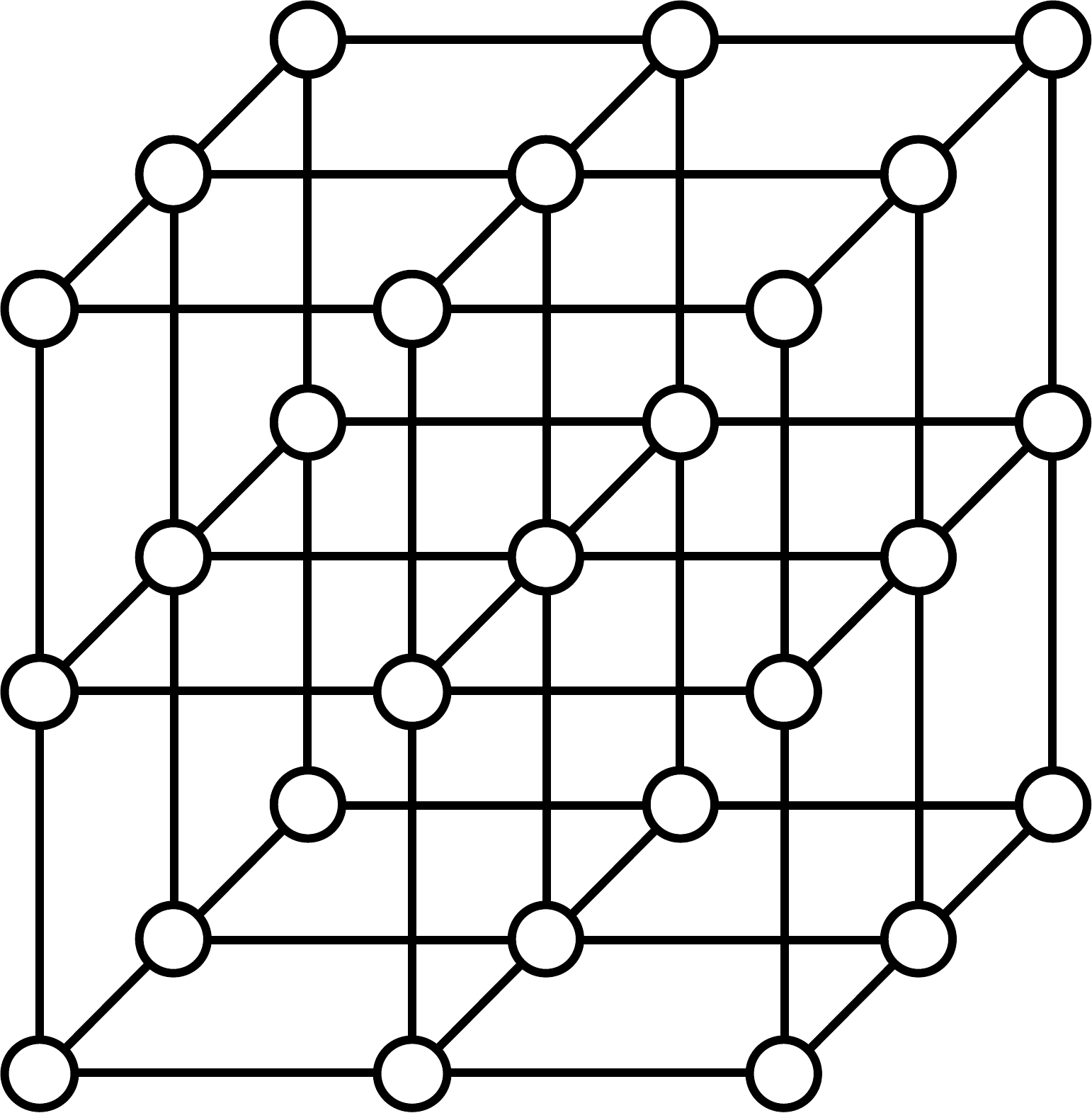}
\vspace{-0.2cm}
\caption{Architecture of the 3D quantum chip.}
\label{fig:3D}
\end{figure}

Besides on the 2D quantum chips, our work advocates optimizing the run-time stack technologies for multi-programming on emerging 3D quantum devices. We simulate a 3D quantum device with 27 physical qubits shown in Figure \ref{fig:3D}. Compared with the 2D quantum chip IBMQ toronto, the qubits in the 3D quantum chip have a higher degree, and the connections between the qubits are denser.

\begin{figure}[t]
\centering
\includegraphics[width=0.99\linewidth]{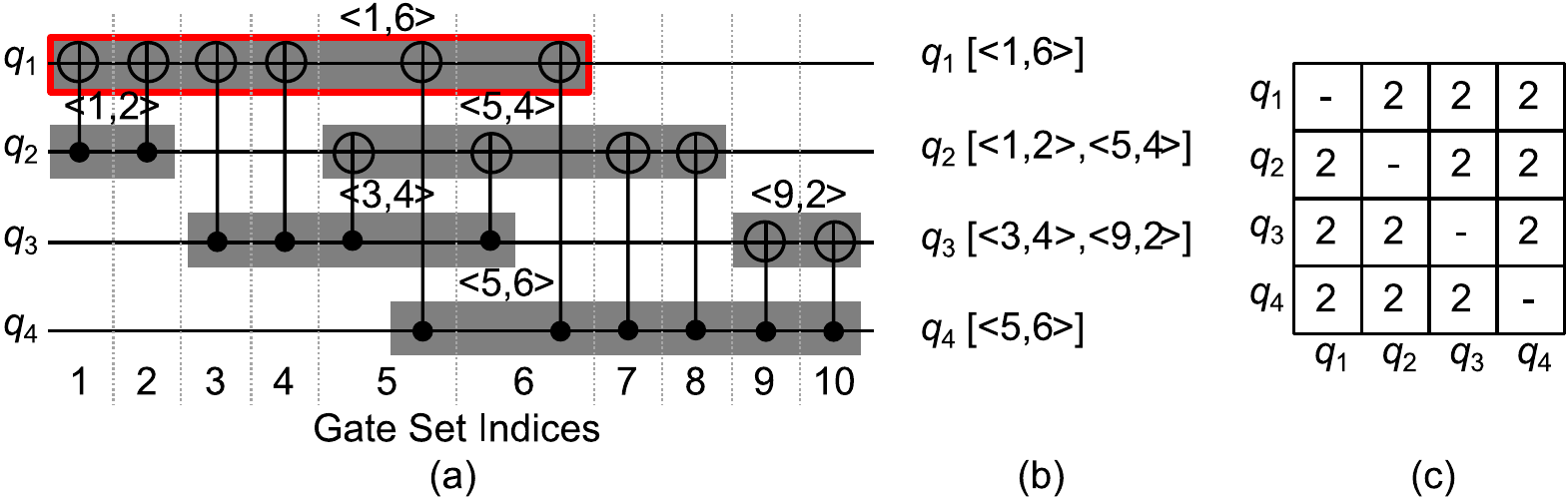}
\vspace{-0.2cm}
\caption{An example for quantum program profiling. (a) The circuit of qft\_4. Single-qubit gates are not shown, as they can be executed without SWAPs. CNOTs that can be executed concurrently can be merged into one gate set. The indices of the gate sets are shown below the circuit. (b) Involvement lists. (c) Coupling strength matrix.}
\label{fig:consecutive}
\end{figure}

\subsection{Quantum Program Profiling\label{part:profiling}}

{\color{black} 
We observe that in a specific quantum circuit, some qubits are consecutively involved in several CNOTs. If a qubit is used in a long sequence of consecutive CNOTs, the qubits it interacts with should be mapped as its neighbors to reduce SWAP overheads. This section shows the opportunity for optimizing qubit mapping by leveraging this observation.

We have a quantum program profiling approach to reveal the CNOT patterns. 1-qubit gates are not considered here as they can be performed without SWAPs. For the example circuit qft\_4 shown in Figure \ref{fig:consecutive}-(a), the profiling outputs the \textit{involvement lists} in Figure \ref{fig:consecutive}-(b). Each program qubit (logic qubit) has a \textit{involvement list} containing several tuples. Each tuple has two values, showing when and for how long the program qubit is involved in consecutive CNOTs. We show these tuples in Figure \ref{fig:consecutive}-(a), and further use dark gray to highlight the stages where qubits are involved in consecutive CNOTs. For instance, the stage highlighted with a red border has a tuple of \textless1, 6\textgreater, indicating that the program qubit $q_1$ is involved in 6 consecutive CNOTs, starting from the gate set with index 1. The profiling algorithm used to extract the \textit{involvement lists} is detailed in Algorithm \ref{algorithm_1}.
Besides, the profiling outputs the \textit{coupling strength matrix} in Figure \ref{fig:consecutive}-(c). The number in each cell denotes the number of interactions (i.e., CNOT gates) between two qubits.

\begin{algorithm}[!t]\fontsize{8}{9}\selectfont
\label{algorithm_1}
\caption{\fontsize{9}{9}\selectfont Involvement list generation.}
\color{black}
\KwIn{Quantum programs}
\KwOut{Involvement\_lists (IL)}
{Initialize involvement\_gates (IG) as an empty list}\;
{Initialize IL as an map, map each program qubit to an empty list}\;

\For{each quantum program}{
    {Gate\_Set\_Index $\gets$ 1}\;
    \For{each gate set of the program (subset if IL is used for initial mapping generation)}{
        {Initialize Gates\_To\_Remove as an empty list}\;
        \For{each gate in IG}{
            \If{any program qubit of this gate is involved in any CNOT in this gate set}{
                {Append the gate to Gates\_To\_Remove}\;
            }
        }
        \For{each gate in the gate set}{
            \For{each program qubit of the gate}{
                {Get the list in IL corresponding to the program qubit}\;
                \eIf{the program qubit is involved in any gate in IG}{
                    {Add 1 to the second item of the last tuple in the list}\;
                }{
                    {Append \textless Gate\_Set\_Index,1\textgreater to the list}\;
                }
            }
        }
        \For{each gate in Gates\_To\_Remove}{
            {Remove the gate from IG};
        }
        \For{each gate in the gate set}{
            {Append the gate to IG}\;
        }
        {Gate\_Set\_Index $\gets$ Gate\_Set\_Index+1}\;
    }
}
\end{algorithm}

These profiling results provide guidance for initial mapping generation and mapping transition. However, the profiling results for the entire circuit are not helpful and can even be misleading for initial mapping generation in practice. Therefore, we extract a subcircuit and use the subcircuit's profiling results for initial mapping generation. We sequentially add each gate set to the subcircuit until each qubit in the program is involved in at least one CNOT gate in the subcircuit, except for ancilla qubits that are not in any CNOTs in the circuit.

For mapping transition, we only use the \textit{involvement list}. Instead of selecting a fixed number of CNOTs for the extended set in topological order for heuristic function, this study selects CNOTs according to the \textit{involvement list}. Refer to Section \ref{part:allocation} and \ref{part:xswap} for more details about how the profiling results are used.
}

\subsection{The Design of a New Qubit Mapping Scheme\label{part:CDAP}}
The initial mapping is critical for a specific quantum program. An excellent initial mapping reduces the SWAP overheads and fully utilizes the robust qubits and links on the quantum chips. Notably, to prevent high crosstalk errors, CNOT gates that cause high crosstalk errors should be avoided in the initial mapping for a specific quantum program and multi-programming cases.

We have the following observations and insights for the quantum chips and multi-programming. (1) The robust qubits and links on a specific quantum chip are limited. (2) Some qubits on the quantum chip have more connections to their surroundings, e.g., as shown in Figure~\ref{fig:2}, $Q_{1}$ has links to the three adjacent physical qubits, while $Q_6$ has a link to only one qubit. (3) Quantum programs have many intra-program qubit interactions and rare inter-program communications. Thus, the qubits in a specific program should be closely allocated (mapped); the allocations for qubits belonging to different quantum programs should avoid mutual interferences, fairly leveraging robust resources. (4) Enabling CNOTs with high crosstalk error possibility should be avoided.

\begin{figure}[t]
\centering
\includegraphics[width=0.99\linewidth]{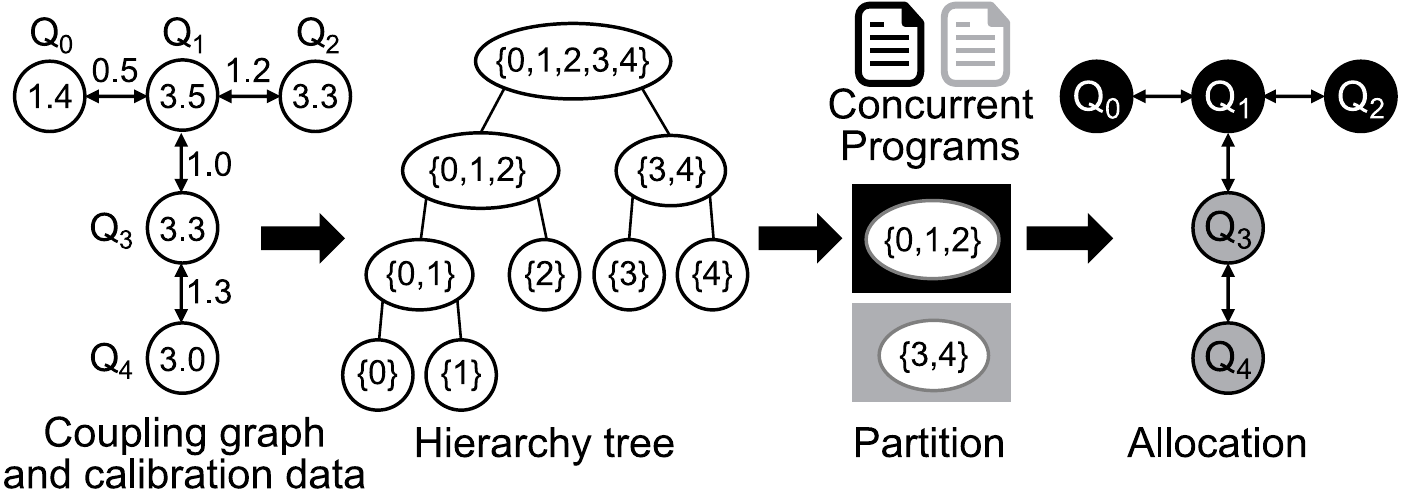}
\vspace{-0.2cm}
\caption{Qubit mapping by using CDAP algorithm.}
\label{fig:7}
\end{figure}

In this paper, we propose a new technology – Community Detection Assistant Partitioning (CDAP) – to construct a hierarchy tree consisted of qubits for searching the robust qubits that are tightly connected for initial allocation. Figure~\ref{fig:7} shows how CDAP works in general. The figure shows the architecture and calibration data of IBM Q London obtained from IBMQ API~\cite{c15_ibmq}. The value in a node represents the readout error rate (in \%) of the qubit, and the value on a link means the error rate (in \%) of the CNOT operation. CDAP creates a hierarchy tree according to the coupling graph and calibration data. In the hierarchy tree, a leaf node denotes a specific physical qubit; an internal node represents the union of its sub-nodes. The values in nodes are the index of the physical qubits. CDAP then partitions by iterating the hierarchy tree from bottom to top to find available regions for each quantum program. Finally, the quantum circuits are allocated to corresponding regions according to qubit degree. We show the details as below.

\subsubsection{Quantum Chip Profiling} The hierarchy tree is a profile of a quantum chip, which helps to locate reliable qubit resources on the quantum computer and avoid high crosstalk CNOT pairs. Algorithm~\ref{algorithm_2} constructs the hierarchy tree based on FN community detection algorithm~\cite{c20_FN}. The algorithm clusters the physical qubits on a specific quantum chip into communities. Qubits in a community have reliable and close interconnections. Couplings in a community have low crosstalk errors. By contrast, the links between communities have relatively low reliability.

\renewcommand{\algorithmcfname}{\fontsize{9}{0}\selectfont Algorithm}
\begin{algorithm}[!t]\fontsize{8}{9}\selectfont
\label{algorithm_2}
\caption{\fontsize{9}{9}\selectfont Hierarchy tree construction.}
\KwIn{Quantum chip coupling graph w/ calibration data}
\KwOut{Hierarchy tree (HT)}
{Initialize HT by setting it empty}\;
{Add a leaf node to HT for each qubit on chip}\;
\While{not all items in HT are merged for a larger community}
{
	{Take two items A and B that are not merged and are with the max value of $F(A, B)$ in HT}\;
	{Create a New\_Node by setting A and B as the New\_Node’s left subtree and right subtree, respectively}\;
	{Append New\_Node to HT}\;
}
{Return HT.}
\end{algorithm}

When the algorithm starts, each physical qubit is a community and is a leaf node in the hierarchy tree. The algorithm keeps merging two communities that can maximize the reward function \emph{F} until there is only one community containing all qubits. Each community during the process corresponds to a node in the hierarchy tree and is a candidate region for allocating qubits. The reward function \emph{F} is defined as the benefit of merging two communities.
\begin{equation}
\setlength{\abovedisplayskip}{3pt}
\setlength{\belowdisplayskip}{3pt}
\begin{aligned}
F=Q_{\text{merged}}-Q_{\text{origin}}+\omega EVX,
\end{aligned}
\label{equ1}
\end{equation} 
in which $Q$ is the modularity of a partition (i.e., $Q=\sum_{i}{(e_{ii}-{a_i}^2)}$~\cite{c20_FN}, in which $e_{ii}$ is the fraction of within-group edges in group $i$, and $a_i$ is the fraction of all edges associated with vertices in group $i$). A higher value of $Q$ indicates a better partition. $Q_\text{origin}$ and $Q_\text{merged}$ denotes the modularity of the original partition and the new partition after merging the two communities, respectively. $E$ denotes the average reliability (i.e., 1 minus the error rate) of CNOTs on the between-group edges of the two communities. $V$ denotes the average reliability of readout operations on the qubits of the two communities.
$X$ denotes the average conditional reliability of CNOTs on the within-group edges that have a crosstalk CNOT in the other community. CDAP takes physical topology, CNOT and crosstalk error rates into account when performing partitioning using the reward function $F$. $\omega$ is a weight parameter. For a specific quantum chip, we can change the value of $\omega$ for adjusting the weight of physical topology and the error rate. If $\omega=0$, CDAP conducts partitioning completely according to physical topology. Noise-awareness is introduced as $\omega$ increases. If $\omega$ keeps increasing, the weight of the error rate will exceed the weight of the physical topology, resulting in the degradation of CDAP to a greedy algorithm that is mainly based on error rate. The mapping results of programs with fewer qubits are more sensitive to $\omega$. Because the variation of $\omega$ changes the error-rate awareness in CDAP, obviously changing the qubit merging order. By contrast, the program with more qubits is less sensitive. More details on how the value of $\omega$ is selected are discussed in Sec. \ref{part:discussions}.

The hierarchy tree has several features: (1) Every node in the hierarchy tree is a candidate region for initial allocation. (2) The physical qubits in a node (i.e., a community) are tightly interconnected. (3) The qubits with a low readout error rate and robust links are preferentially merged. Thus, the more robust the qubit set is, the higher the node depth will be. Whether the hierarchy tree is balanced or not doesn't impact the qubit mapping result, as the most reliable region can always be selected for a specific program. The hierarchy tree helps to locate the robust resources on quantum computers, providing better initial mapping for quantum programs.

Further, we explain why the hierarchy tree helps to select the initial allocation with the example shown in Figure~\ref{fig:7}. (\romannumeral1) $Q_0$ and $Q_1$ are firstly merged due to the link between them is with the lowest error rate. (\romannumeral2) Then, $Q_2$ instead of $Q_3$ is merged into the community \{0, 1\}, though the link $Q_1$-$Q_3$ has a lower CNOT error rate than $Q_1$-$Q_2$. This is because the algorithm tends to merge more interconnected nodes into one community, avoiding the waste of robust physical qubits. Likewise, $Q_3$ and $Q_4$ are merged. (\romannumeral3) Finally, all qubits are merged as the root of the hierarchy tree. The algorithm avoids wasting robust resources caused by the topology-unaware greedy algorithm and supports more quantum programs to be mapped on a specific quantum chip.

As the calibration data doesn't change frequently (e.g., IBM calibrates the devices once a day~\cite{c15_ibmq}), the hierarchy tree only needs to be constructed once in each calibration cycle. It can be saved for possible reuse within 24 hours, without incurring more overheads.

\begin{algorithm}[!t]\fontsize{8}{9}\selectfont
\label{algorithm_3}
\caption{\fontsize{9}{9}\selectfont Qubit partition.}
\KwIn{Hierarchy tree, Quantum programs}
\KwOut{Partition}
\If{there is only one quantum prorgam to be mapped}{
    {Add the root of the Hierarchy tree to Partition and return Partition}\;
}
{Sort quantum programs in descending order of \emph{CNOT density}}\;
\For{each quantum program}{
	{Initialize candidate nodes (C) as an empty set}\;
	\For{each leaf node in the hierarchy tree}{
    	{Search a node that has enough number of on-chip qubits for allocating the program from the leaf to its parent nodes}\;
    	{Add the node to C}\;
	}
	
	\If(~~\#~The current qubit state cannot meet the requirements){C is empty}{
		{Fail and revert to separate execution}\;
	}
	{Construct a tuple for each node in C for sorting (detailed in section \ref{sec:partition})}\;
	{Add the the first node in sorted C to Partition}\;
	{Remove qubits in the node from all other nodes in the hierarchy tree}\;
	\If(~~\#~The sibling node has no path to other nodes in the hierarchy tree){the node's sibling node is isolated}{
		{Remove qubits in the sibling node from its parent nodes}\;
		{Remove the link from the sibling node to its parent}\;
	}
}
{Return Partition.}
\end{algorithm}

\subsubsection{Qubit Partitioning for Concurrent Circuits \label{sec:partition}} For multi-programming cases, Algorithm~\ref{algorithm_3} partitions the qubits into multiple regions for concurrent programs according to the hierarchy tree. The algorithm prioritizes the programs with a higher value of the \emph{CNOT density}, which is defined as (the number of CNOT instructions) / (the number of program qubits). For each program, the algorithm searches the hierarchy tree from bottom to top to find the candidate qubit sets available for the program.

{\color{black}
Selecting the qubit set for a specific quantum program is based on the following principles. (1) Candidates with dense coupling can reduce additional SWAPs. (2) Candidates with fewer physical qubits can save resources. (3) Candidates with more reliable qubits can improve fidelity. (4) Candidates with less crosstalk-prone CNOT pairs lead to fewer barriers and lower decoherence errors. 

We construct a tuple for each candidate for sorting. The tuple has four items. For the first item, we get all possible mappings for the quantum program on this qubit set candidate. Then, we have the average of the shortest path lengths between any two program qubits for each possible mapping; the smallest average value is used as the first item in the tuple. The second is the number of physical qubits in the candidate. The third is the average error rate of the candidate. For the fourth item, we get all possible CNOT pairs and find the crosstalk-prone ones (CNOT pairs with an error amplification ratio greater than 2 are crosstalk-prone); then, the number of crosstalk-prone CNOT pairs is used as the fourth item in the tuple. Finally, we sort the candidates in ascending order based on the tuple. The first one in the sorted candidates is selected as the most appropriate qubit set for the quantum program.}

\begin{algorithm}[!t]\fontsize{8}{9}\selectfont
\label{algorithm_4}
\caption{\fontsize{9}{9}\selectfont Qubit allocation.}
\KwIn{Quantum program, Involvement lists, Coupling strength matrix}
\KwOut{Layout}
{Get the subgraph containing all physical qubits assigned to the quantum program}\;

{Get the degree of each physical qubit in the subgraph}\;

{Get the degree of each program qubit according to the coupling strength matrix}\;

{Initialize the list of pending program qubits (P)}\;

{Sort P based on the first tuple in each program qubit's involvement list}\;

{Initialize candidate qubits (C) as a list containing all physical qubits whose degree is greater than or equal to the degree of the first pending program qubit in P}\;

\eIf{C is not empty}{
    {Select the physical qubit with the lowest average CNOT error rate in C}\;
}{
    {Select the physical qubit with the lowest value of 1/(degree $\times$ average CNOT reliablity) in the subgraph}\;
}
{Map the first pending program qubit in P to the selected physical qubit}\;

{Remove the first pending program qubit from P}\;

\For{each pending program qubit in P}{
    {Get neighbors of mapped physical qubits}\;
    
    {Select the neighbor that provides the most reliable SWAP paths between this program qubit and all other program qubits that have been mapped}\;
    
    {Map the pending program qubit to the selected neighbor}\;
    
    {Remove the pending program qubit from P}\;
}
\end{algorithm}

\subsubsection{Qubit Allocation Leveraging High-degree Qubits \label{part:allocation}} SWAP operations are required when two program qubits are not mapped adjacently \cite{c52_qucloud,c71_PACT}. Allocating the program qubit with more interactive neighbors to a physical qubit with a higher degree is effective to reduce the SWAP overheads, especially for the future quantum chips with a denser structure and higher dimension (e.g., 3D quantum chips). We devise a new qubit mapping strategy that utilizes high-degree physical qubits to enhance initial mapping. Details are in Algorithm \ref{algorithm_4}. First, we have the degree of physical qubits and program qubits. We refer to the degree of a physical qubit as the number of its connections to adjacent physical qubits. The degree of a program qubit is the number of program qubits that interact with it. Next, all qubits of the quantum program are added to the list of pending program qubits. We sort the pending program qubits in ascending order of the first tuple's first value in each qubit's \textit{involvement list}. Then, we select a high-degree physical qubit to map the first pending program qubit. Details are in lines 6 to 13 in Algorithm \ref{algorithm_4}. Finally, for each of the rest pending program qubits, we select the most appropriate neighbor of mapped physical qubits. We map the program qubit onto the physical qubit that provides the most reliable SWAP paths between this program qubit.

\subsubsection{Discussion\label{part:discussions}} Our design merges the reliable qubits with robust links and the lower read-out error rate into a specific community in each iteration. Unreliable qubits would be added into the community at last. When performing qubits allocation, CDAP searches the hierarchy tree from bottom to top to find candidates for partition. Unreliable qubits are less likely to be selected, thereby improving the overall fidelity.

Using CDAP might lead to a case where the allocated qubits for a quantum program exceed the qubits that the program needs. For example, if a 4-qubit quantum program is mapped on the quantum chip in Figure~\ref{fig:7}, the only available region (community) is the root of the hierarchy tree, i.e., \{0,1,2,3,4\}, leaving one qubit unmapped/unused, i.e., the redundant qubit. To avoid waste, we label these redundant qubits and add them to adjacent communities.

\begin{figure}[t]
\centering
\includegraphics[width=0.80\linewidth]{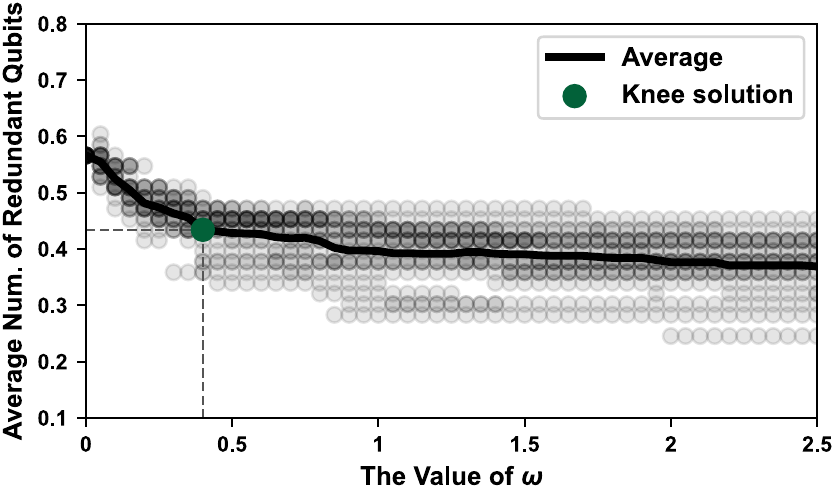}
\vspace{-0.2cm}
\caption{The average number of redundant qubits in the hierarchy tree varies with $\omega$. A gray dot with darker color represents more cases are overlapping in this result. The knee solution refers to the value of $\omega$, which makes the change of redundant qubits slow down with the increase of $\omega$. We use the knee solution because it can reduce the redundant qubits as much as possible, without making the community partitioning depend too much on error rate.}
\label{fig:9}
\end{figure}

For a specific node in the hierarchy tree, we define the term maximum redundant qubits, which refers to the maximum possible number of unused qubits when a quantum program is allocated to the community. The number of maximum redundant qubits of a node is:  \emph{$\text{node}.\text{n\_qubits}-(1+\max(\text{node}.\text{left}.\text{n\_qubits},   \text{node}.\text{right}.\text{n\_qubits}))$}. We observe that the increase of $\omega$ in the reward function leads to the degradation of the hierarchy tree, i.e., in each merge process when constructing a hierarchy tree, only one leaf node containing one qubit is added to the new community. The number of maximum redundant qubits of the new community is 0 in this case. Thus, the increase of $\omega$ leads to a reduction in average redundant qubits. As illustrated in Figure~\ref{fig:9}, we vary the $\omega$ from 0 to 2.5 for 20 different calibration data of IBMQ toronto, and record the average number of the redundant qubits in the hierarchy tree. We take the knee solution, in which the value of $\omega$ is 0.4. In this case, CDAP takes both physical topology and the error rate into account. 

\subsection{The Design of The Mapping Transition Scheme}

Multi-programming brings new challenges for mapping transition. In this paper, we design the X-SWAP, which includes both inter- and intra-program SWAP operations. It provides the ideal SWAP solution by considering the SWAP possibilities across all of the qubits. In practice, the inter-program SWAP can be enabled when two quantum programs are close to each other. The cost of inter-program SWAPs can be less than the cost in the cases where only intra-program SWAPs are used. The below section shows the details.

\begin{figure}[t]
\centering
\includegraphics[width=0.77\linewidth]{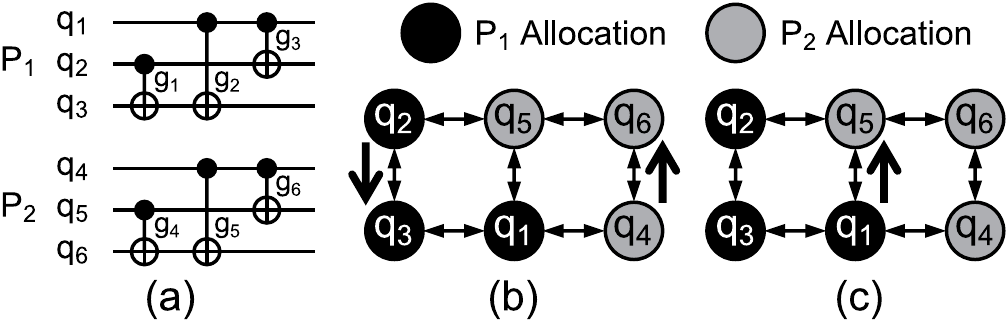}
\vspace{-0.2cm}
\caption{A case for an inter-program SWAP.}
\label{fig:6}
\end{figure}

\subsubsection{The Advantages of Inter-program SWAP}
We find two main advantages of inter-program SWAP. First, an inter-program SWAP can replace two or more intra-program SWAPs. Fig \ref{fig:6} shows a case where two quantum programs are mapped on a specific quantum chip. Figure \ref{fig:6}-(a) shows the two quantum programs ($P_1$ and $P_2$), and (b) illustrates how they are mapped on a quantum chip with 6 physical qubits. According to the initial mapping, for $P_1$, the $g_1$ and $g_2$ can be executed directly – without any SWAP operations involved. However, the $g_3$ cannot be executed directly unless a SWAP operation between $q_2$ and $q_3$ is executed first. Same thing happens for $P_2$. A SWAP operation between $q_4$ and $q_6$ should be involved before $g_6$ can be executed. To sum up, for such a mapping case where two programs are involved, two SWAPs are needed. By contrast, if the two programs could be mapped together, the inter-program SWAP operation could be enabled. Figure \ref{fig:6}-(c) shows a new policy that enables inter-program SWAPs, where only one inter-program SWAP \{$q_1$, $q_5$\} is needed. The intra-program SWAPs, i.e., \{$q_2$, $q_3$\}, \{$q_4$, $q_6$\}, can be replaced by one inter-program SWAP across $q_1$ and $q_5$. Obviously, using the inter-program SWAP achieves the same goal but has lower overheads.

\begin{figure}[t]
\centering
\includegraphics[width=0.77\linewidth]{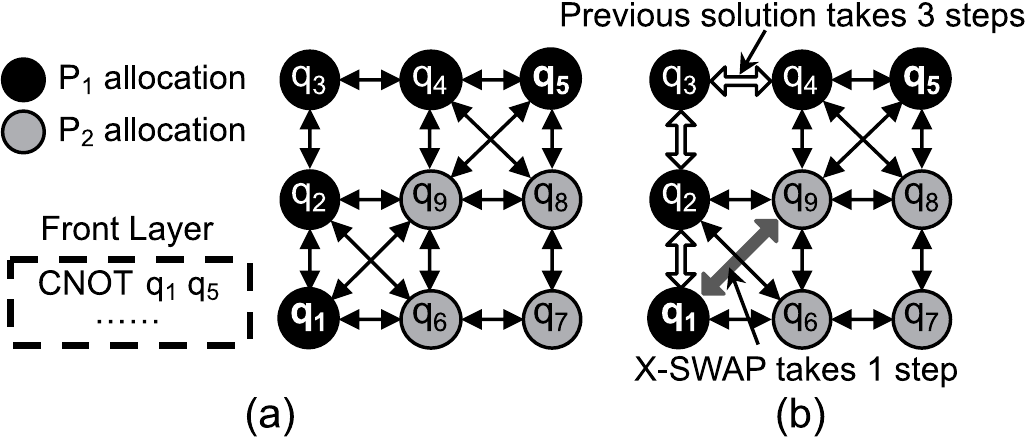}
\vspace{-0.2cm}
\caption{(a) $P_1$ and $P_2$ are mapped on a quantum chip with 9 qubits. The next gate to be solved is CNOT that involves $q_1$ and $q_5$. (b) X-SWAP scheme takes shortcuts to satisfy the constraint of CNOT $q_1$, $q_5$.}
\label{fig:10}
\end{figure}

Second, Inter-program SWAPs take shortcuts. For instance, Figure~\ref{fig:10}-(a) shows two quantum programs are co-located (mapped) on a quantum chip with nine qubits. $q_1$ and $q_5$ are not mapped physically adjacent; SWAPs are required to satisfy their constraint to make CNOT $q_1$, $q_5$ executable. As illustrated in Figure \ref{fig:10}-(b), an inter-program SWAP i.e., \{$q_1$, $q_9$\}, takes only one step (swap operation) to move $q_1$ and $q_5$ adjacent. By contrast, to achieve the same goal, previous intra-program scheme has to introduce three SWAPs, i.e., \{$q_1$, $q_2$\}, \{$q_1$, $q_3$\}, \{$q_1$, $q_4$\}. Briefly, enabling inter-program SWAPs could result in fewer SWAPs in the cases where multiple quantum programs are mapped as neighbors on a specific quantum chip, therefore reducing the SWAP overheads and benefiting the overall fidelity.

\begin{figure}[t]
	\centering
	\begin{minipage}[t]{0.40\linewidth}
		\center
		\includegraphics[width=0.75\linewidth]{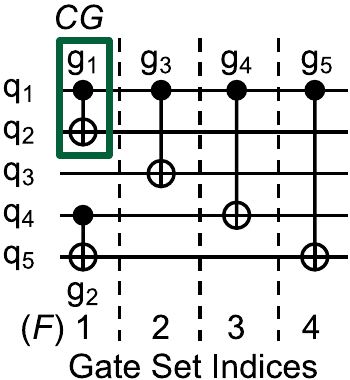}
		\vspace{-0.2cm}
		\caption{Critical gates.\label{fig:11}}
	\end{minipage}
	\begin{minipage}[t]{0.58\linewidth}
		\center
		\includegraphics[width=0.99\linewidth]{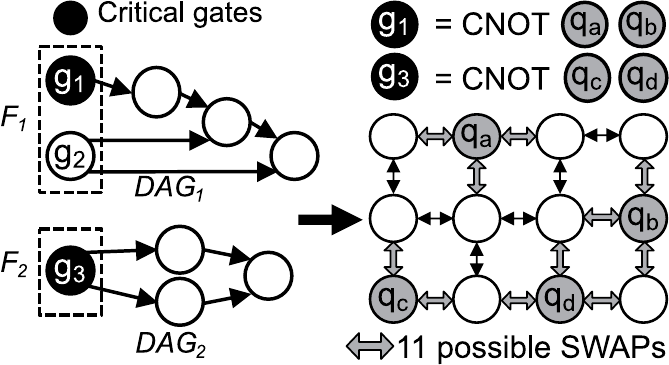}
		\vspace{-0.61cm}
		\caption{Example of SWAP candidates.}\label{fig12}
	\end{minipage}
\end{figure}

\subsubsection{X-SWAP\label{part:xswap}}
Instead of generating a schedule for each quantum program separately and then merge them, which is done by the previous work, we are the first to design an approach for generating the global scheduling solution for all of the programs simultaneously. In our work, the SWAP-based heuristic search scheme in previous work~\cite{c12_SABRE} is used as the baseline. The design details are shown as follows.

\noindent\textbf{Heuristic search space.} To show our design, we illustrate a circuit for a quantum program $P$ in Figure~\ref{fig:11}. The CNOT gates of the circuit can be divided into 4 gate sets. CNOTs in a gate set can be executed in parallel. The first gate set is the Front layer (denoted as $F$) of $P$, which denotes the set of all gates without unexecuted predecessors in the Directed Acyclic Graph (\emph{DAG}) of $P$. Critical Gates (\emph{CG}) denote the set of CNOT gates in $F$ that have successors on the second gate set. For example, in $F$, $g_1$ has a successor $g_3$ on the second gate set, but $g_2$ has no successors. Thus, $g_1$ is a critical gate; $g_2$ is not. If the critical gate $g_1$ is executed and removed from the \emph{DAG}, the data dependency of $g_3$ is resolved and the front layer $F$ will be updated. By contrast, handling $g_2$ firstly doesn't help to update $F$.

For each program $P_i$, we remove hardware-compliant gates that can be executed directly in $F_i$ from $\text{\emph{DAG}}_i$, and apply them to the mapped circuit. When there are no hardware-compliant gates, SWAPs are needed to make hardware-incompliant gates executable. Among all of the hardware-incompliant gates, the data dependency of critical gates need to be handled firstly for the purpose of updating the Front Layer and reducing the post-mapping circuit depth. Thus, we only search the SWAPs associated with qubits in critical CNOT gates. To help understand, Figure \ref{fig12} shows another example, in which $g_1$ and $g_3$ are critical gates illustrated in DAGs. The qubits involved in $g_1$ and $g_3$ are not mapped as the close neighbors. All SWAPs associated with the critical gates are SWAP candidates. They are highlighted on the coupling map of the quantum chip. The best SWAP among candidates is selected according to the heuristic cost function (details refer to the following). The mapping is updated as the SWAP is applied to the mapped circuit. Some hardware-incompliant CNOTs become executable when their constraints are eliminated using SWAPs. This repeats until the constraints of all CNOTs in the DAG are satisfied. 

\begin{algorithm}[t]\fontsize{8}{9}\selectfont
\label{algorithm_5}
\caption{\fontsize{9}{9}\selectfont X-SWAP mechanism.}
\KwIn{Quantum chip coupling graph, Quantum programs, Initial mapping}
\KwOut{Final Schedule (FS)}
{Generate a DAG for each program}\;
{Generate a Front Layer for each DAG}\;
\While{not all gates' constraints are satisfied}{
	\eIf{hardware-compliant gates exist}{
		{Append hardware-compliant gates to FS}\;
		{Remove hardware-compliant gates from the DAG and update the Front Layer}\;
	}{
		\For{each Front Layer}{
			{Append CNOTs in the Front Layer that have subsequent CNOTs on the second layer to Critical Gates}\;
		}
		{Add SWAPs associated with the qubits in Critical Gates to SWAP candidates}\; 
		{Find a SWAP with the lowest crosstalk error from the SWAP candidates and with the minimum value of \emph{score}(\emph{SWAP})}\;
		{Append the SWAP to FS and update mapping}\;
	}
}
{Return FS.}
\end{algorithm}

\noindent\textbf{Design of the heuristic cost function.} The heuristic cost function gets the best SWAP from all SWAP candidates (inter/intra-program SWAPs). We show its core idea.

The concept \emph{locality} for mapping a quantum program is critical. It indicates that the mapping policy should keep qubits belonging to a specific program close to each other. Otherwise, high SWAP overheads will occur between two qubits mapped far away from each other when required for a CNOT operation. Nearest Neighbor Cost (NNC) is the length of the shortest path between two program qubits mapped on a quantum chip. NNC-based heuristic function (\emph{H}) is used in SABRE~\cite{c12_SABRE} to choose the best SWAP among the SWAP candidates. We also use it as a component in our approach. It represents the sum of the cost in the front layers and the cost in the extended sets \cite{c12_SABRE}. The extended set contains a fixed number of closest successors of the gates in the front layer. Each set's cost is calculated as the averaged NNC of all CNOT gates in the set.

We optimize \emph{H} by selecting gates for the extended set based on the \textit{involvement list}. The cost of extended set is introduced in SABRE \cite{c12_SABRE} for look-ahead ability. However, the look-ahead ability is limited because the size of the extended set is fixed. The gates in the extended set are added in a topological order. Thus, not all gates that contribute to a proper SWAP can be added to the extended set. Some gates added into the extended set can even be misleading, resulting in SWAPs that would incur higher SWAP overheads. To tackle this problems, we select gates for the extended set according to the \textit{involvement list}. When all gates in the front layer are not executable, we initialize extended gate as an empty set. We first get the gate set index of each gate in the front layer. For each program qubit, we get the last tuple from its \textit{involvement list} that satisfies the constraint - the tuple's first item is less equal than any gate set index of gates in the front layer. A tuple means a specific program qubit involved in consecutive CNOTs. For each CNOT in the set, if it is neither applied to the mapped circuit nor in the front layer, it is added to the extended set.

We also prioritize inter-program SWAPs on the shortest SWAP path. Given the coupling map of a quantum chip and the qubit allocations, we define the term distance matrix $ D $, in which each cell denotes the length of the shortest path between two physical qubits on the quantum chip. For each program $P_i$, we define $D_{i}^{'}$ as the shortest path matrix for qubits that have not been occupied by other programs. i.e., unmapped physical qubits and the physical qubits on which $P_i$ is mapped. In essence, $D$ represents the shortest path matrix to perform mapping transition for concurrent quantum programs with inter-program SWAPs enabled; $D_{i}^{'}$ represents the shortest path matrix to perform mapping transition for $P_i$. For a two-qubit gate $g$, we denote the two program qubits involved in $g$ as $g.q_1$ and $g.q_2$. We define the physical qubit on which a program qubit $q$ is mapped as $\sigma(q)$. The shortest path between two qubits involved in a two-qubit gate minus 1 is the minimum number of SWAPs required to satisfy their constraint.

In our design, if $D_{i}^{'}[\sigma(g.q_1)][\sigma(g.q_2)]$ is greater than $D[\sigma(g.q_1)][\sigma(g.q_2)]$ for a two-qubit gate $g$ in a specific quantum program $P_i$, it means that inter-program SWAPs outperforms intra-program SWAPs when satisfying the constraint of $g$. In such cases, the X-SWAP scheme should enable inter-program SWAPs to reduce the quantum programs' mapping transition cost. For example, in Figure~\ref{fig:10}-(b), as it takes either 1 inter-program SWAP or 3 intra-program SWAPs to satisfy the constraint of CNOT $q_1$, $q_5$, it delivers $D[\sigma(q_1)][\sigma(q_5)]=2$ and $D_{1}^{'}[\sigma(q_1)][\sigma(q_5)]=4$. In terms of inserting SWAPs to satisfy $g$'s constraint, we define the number of SWAPs saved by X-SWAP scheme as below:
\begin{equation} 
\setlength{\abovedisplayskip}{3pt}
\setlength{\belowdisplayskip}{3pt}
gain(g)=D[\sigma(g.q_1)][\sigma(g.q_2)]-D_i^{'}[\sigma(g.q_1)][\sigma(g.q_2)],
\end{equation}
and, we define the heuristic cost function as below:
\vspace{-0.01cm}
\begin{equation}
\setlength{\abovedisplayskip}{3pt}
\setlength{\belowdisplayskip}{3pt}
\begin{aligned}
score(\text{\emph{SWAP}})=&H(\text{\emph{SWAP}})+\\
&\sum_{F_i\in F}{\frac{1}{|F_i|}\sum_{g\in F_i}{gain(g)I(\text{\emph{SWAP}},g)}}.\\
\end{aligned}
\end{equation}

As the sizes of different Front Layers vary, the gain is normalized to their sizes accordingly. The shortest SWAP path for satisfying $g$'s constraint involves several qubits. When both program qubits involved in the \emph{SWAP} is on the shortest SWAP path, $I(\text{\emph{SWAP}},g) = 1$. Otherwise, $I(\text{\emph{SWAP}},g) = 0$. This indicates only the SWAPs on the shortest SWAP path are prioritized. The SWAP with the minimum value of $score$ is the best among the candidates. When there are more than one SWAPs with the minimum score, we select the SWAP with the lowest crosstalk error to mitigate crosstalk. Algorithm~\ref{algorithm_5} shows the overall logic of X-SWAP.

\subsection{The Design of the QuCloud+ Scheduler}
{\color{black} Randomly selected quantum programs for multi-programming workloads often lead to qubit resource under-utilization or fidelity degradation. We design the QuCloud+ scheduler to select appropriate combinations of concurrent quantum programs that promise high fidelity and high throughput simultaneously.} Algorithm \ref{algorithm_6} shows its core scheduling logic.

The first step is to generate co-location candidates. The scheduler selects candidates that can be co-located on the quantum chip with the first incoming job $P_0$. The total number of qubits in the selected workloads cannot exceed that of the quantum chip. The selected candidates are sorted in descending order of the $fitness$ with $P_0$. Quantum programs with fewer qubits require fewer resources and help to protect fidelity when robust resources are limited. Moreover, quantum programs with similar circuit depths are unlikely to cause serious decoherence errors in multi-programming cases. Because the deeper quantum circuit needs more execution time and thus other co-located circuit may suffer decoherence in waiting for the deeper circuit completion. Thus, the $fitness$ between $P_0$ and another quantum program $P_i$ in candidates is calculated as follows:
\begin{equation}
\setlength{\abovedisplayskip}{3pt}
\setlength{\belowdisplayskip}{3pt}
fitness=1/(P_i.n\_qubits\times S(|P_i.depth-P_0.depth|)),
\end{equation}
where $S(x)$ denotes the Sigmoid function, i.e., $S(x)=\frac{1}{1+e^{-x}}$. The function can map $x$ in the interval $[0,+\infty]$ to the interval $[0.5,1]$. The greater the difference between the depths of $P_0$ and $P_i$, the closer the value of $S(x)$ is to 1. By contrast, the smaller the difference is, the closer the value of $S(x)$ is to 0.5, i.e., the $fitness$ is doubled.

\begin{algorithm}[tbp]\small
\label{algorithm_6}
\caption{\fontsize{9}{9}\selectfont The QuCloud+ scheduler.}
\KwIn{List of incoming jobs (IJ), Hierarchy tree}
\While{not all jobs in IJ are scheduled}{
	{Initialize cur\_job\_set as a list having the first job in IJ}\;
	{Initialize candidates as a list containing the following $N$ jobs that can be co-located with the first job}\;
    {Sort candidates in descending order of the fitness with the first job}\;
	\For{each candidate in candidates}{
		{Append candidate to cur\_job\_set}\; 
		\For{each job in cur\_job\_set}{
			{Calculate job's sep\_EPST}\;
			{Calculate job's co\_EPST}\;
			{Calculate job's EPST\_violation as 1-(co\_EPST/ sep\_EPST)}\;
			\If{EPST\_violation $>$ $\varepsilon$}{
				{Remove candidate from cur\_job\_set}\;
				{Break}\;
			}
		}
		\If{cur\_job\_set's length = $M$}{
		    {Break}\;
		}
	}
	{Algorithm~\ref{algorithm_5} is called to map programs in cur\_job\_set}\;
	{Submit cur\_job\_set for execution}\;
	{Remove all programs in cur\_job\_set from IJ.}
} 
\end{algorithm}

{\color{black} The second step is threshold-based workload selection. The Estimated Probability of a Successful Trial (EPST) is proposed to estimate the fidelity of the execution of a quantum program on a specific quantum chip. It is defined as bellow:
\begin{equation}
\setlength{\abovedisplayskip}{3pt}
\setlength{\belowdisplayskip}{3pt}
EPST=r_{2q}^{|\text{\emph{CNOTS}}|} r_{1q}^{|\text{\emph{1q gates}}|} r_{ro}^{|\text{\emph{qubits}}|},
\end{equation}
in which $r_{2q}$, $r_{1q}$ and $r_{ro}$ denotes the average reliability of CNOTs, the average reliability of 1-qubit gates, and the average reliability of readout operations on the allocated physical qubits, respectively. $|\text{\emph{CNOTS}}|$, $|\text{\emph{1q gates}}|$ and $|\text{\emph{qubits}}|$ denotes the number of CNOT gates, the number of single-qubit gates, and the number of qubits of the quantum program, respectively. A higher EPST indicates the quantum program is mapped to a region with more robust resources and also indicates a higher PST may be obtained during the real execution in practice. 

Separate EPST (Sep\_EPST) is the maximum EPST that a program can achieve. To obtain the Sep\_EPST, Algorithm~\ref{algorithm_3} is called to allocate a set of physical qubits for every single quantum program. Co-located EPST (Co\_EPST) represents the EPST when multiple quantum programs are co-located on a quantum chip. To obtain the Co\_EPST, Algorithm~\ref{algorithm_3} is called to generate a partition for all programs in the workload.} EPST violation is calculated according to Sep\_EPST and Co\_EPST of each quantum program. If the EPST violation of all quantum programs in the workload is less than the threshold $\varepsilon$, the workload can be co-located on the chip. Otherwise, they can not be co-located. 

QuCloud+ scheduler supports to co-locate more than two quantum programs on a quantum computer. We set the maximum number of co-located programs to $M$ (3 by default) to avoid severe performance degradation. To ensure efficiency and fairness, the size of the candidates is set to $N$ (10 by default in practice). More details are in Algorithm~\ref{algorithm_6}.

\section{Evaluations}
\subsection{Methodology}
\subsubsection{Metrics} The following metrics are used for evaluations.

\noindent\textbf{Probability of a Successful Trial (PST).} PST is used to evaluate the fidelity of the quantum program execution~\cite{c11_not_all,c8_multi,c29_mitigate}. PST is defined as the fraction of trails that produce a correct result. To get PST, we map and run each workload on the target quantum chip for 8192 trials.

\noindent\textbf{Number of post-mapping CNOT gates.} We use the number of post-mapping CNOT gates to evaluate the policy's ability to reduce the SWAP overheads when mapping multiple quantum programs.

\noindent\textbf{Post-mapping circuit depth.} The quantum program's post-mapping circuit depth is used to evaluate the policy's capability for reducing the coherence error.

\noindent\textbf{Trial Reduction Factor (TRF).} TRF is used to evaluate the improvement of the throughput brought by multi-programming policies~\cite{c8_multi}. TRF is defined as the ratio of needed trails when programs are executed separately to the trails needed when multi-programming is enabled.

\subsubsection{Simulators}
For 2D quantum chips, we simulate IBMQ Toronto~\cite{c15_ibmq} and IBMQ50~\cite{c27_ibmq50} using the QASM simulator of Qiskit\cite{c13_qiskit}. To reveal the architectural characteristics of future 3D quantum devices, we also construct a simulated 3D quantum chip. Its architecture is shown in Figure \ref{fig:3D}. As the QASM simulator has no support for conditional error rate, we follow the work in \cite{c60_crosstalk} to simulate the conditional error rate characterized by SRB. We use the realistic topology information and calibration data of IBMQ Toronto and randomly generate the crosstalk amplification ratios. We further generate calibration data for IBMQ50 and the 3D quantum chip using a uniform random model.

\begin{scriptsize}
\begin{table}[t]
    \caption{NISQ Benchmarks.}
    \vspace{-6pt}
  \centering
  \setlength{\tabcolsep}{3.5pt}
    \begin{tabular}{|c|c|l|c|c|l|}
    \hline
    \multicolumn{6}{|c|}{\textbf{Benchmarks}} \\
    \hline
    \textbf{type} & \textbf{ID} & \multicolumn{1}{c|}{\textbf{Name}} & \textbf{type} & \textbf{ID} & \multicolumn{1}{c|}{\textbf{Name}} \\
    \hline
    \hline
    \multirow{6}*{tiny-sized} & 1     & bv\_n3 & \multirow{14}*{large-sized} & 15    & 4gt4-v0\_72 \\
\cline{2-3}\cline{5-6}          & 2     & bv\_n4 &       & 16    & sf\_276 \\
\cline{2-3}\cline{5-6}          & 3     & peres\_3 &       & 17    & alu-bdd\_288 \\
\cline{2-3}\cline{5-6}          & 4     & toffoli\_3 &       & 18    & ex2\_227 \\
\cline{2-3}\cline{5-6}          & 5     & fredkin\_3 &       & 19    & ham7\_104 \\
\cline{2-3}\cline{5-6}          & 6     & xor5\_254 &       & 20    & C17\_204 \\
\cline{1-3}\cline{5-6}    \multirow{6}*{small-sized} & 7     & 3\_17\_13 &       & 21    & bv\_n10 \\
\cline{2-3}\cline{5-6}          & 8     & 4mod5-v1\_22 &       & 22    & ising\_model\_10 \\
\cline{2-3}\cline{5-6}          & 9     & mod5mils\_65 &       & 23    & qft\_10 \\
\cline{2-3}\cline{5-6}          & 10    & alu-v0\_27 &       & 24    & sys6-v0\_111 \\
\cline{2-3}\cline{5-6}          & 11    & decod24-v2\_43 &       & 25    & sym9\_146 \\
\cline{2-3}\cline{5-6}          & 12    & 4gt13\_92 &       & 26    & rd53\_311 \\
\cline{1-3}\cline{5-6}    \multirow{2}*{large-sized} & 13    & aj-e11\_165 &       & 27    & qft\_16 \\
\cline{2-3}\cline{5-6}          & 14    & alu-v2\_31 &       & 28    & cnt3-5\_180 \\
    \hline
    \end{tabular}%
  \label{tbl1}
\end{table}
\end{scriptsize}

\begin{scriptsize}
\begin{table*}[tp]
  \caption{PST comparison between multiple strategies on IBMQ toronto.}
  \vspace{-6pt}
  \centering
  \setlength{\tabcolsep}{1.6pt}
    \begin{tabular}{|c|c|c|c|c|c|c|c|c|c|c|c|c|c|c|c|c|c|c|c|c|c|c|}
    \hline
    \multicolumn{2}{|c|}{\textbf{Workloads}} & \multicolumn{3}{c|}{\textbf{Separate}} & \multicolumn{3}{c|}{\textbf{SABRE}} & \multicolumn{3}{c|}{\textbf{Baseline}} & \multicolumn{3}{c|}{\textbf{QuCloud}} & \multicolumn{3}{c|}{\textbf{QuCloud+}} & \multicolumn{3}{c|}{\textbf{CDAP-only}} & \multicolumn{3}{c|}{\textbf{X-SWAP-only}} \\
    \hline
    \textbf{ID1} & \textbf{ID2} & \textbf{PST1} & \textbf{PST2} & \textbf{avg} & \textbf{PST1} & \textbf{PST2} & \textbf{avg} & \textbf{PST1} & \textbf{PST2} & \textbf{avg} & \textbf{PST1} & \textbf{PST2} & \textbf{avg} & \textbf{PST1} & \textbf{PST2} & \textbf{avg} & \textbf{PST1} & \textbf{PST2} & \textbf{avg} & \textbf{PST1} & \textbf{PST2} & \textbf{avg} \\
    \hline
    \hline
    1     & 1     & 90.50  & 89.14  & 89.82  & 65.97  & 79.24  & 72.60  & 80.41  & 67.08  & 73.74  & 88.39  & 88.90  & 88.65  & 89.18  & 89.45  & 89.32  & 89.53  & 89.43  & 89.48  & 78.71  & 56.46  & 67.58  \\
    \hline
    1     & 2     & 89.51  & 85.31  & 87.41  & 68.75  & 62.79  & 65.77  & 81.65  & 54.24  & 67.94  & 88.49  & 76.55  & 82.52  & 88.42  & 75.20  & 81.81  & 89.15  & 74.37  & 81.76  & 52.99  & 54.46  & 53.72  \\
    \hline
    1     & 3     & 90.37  & 57.14  & 73.75  & 72.17  & 87.61  & 79.89  & 81.51  & 78.61  & 80.06  & 88.27  & 84.63  & 86.45  & 88.79  & 90.04  & 89.42  & 88.28  & 90.01  & 89.15  & 77.66  & 81.45  & 79.55  \\
    \hline
    1     & 4     & 89.48  & 58.02  & 73.75  & 79.39  & 61.13  & 70.26  & 81.64  & 85.19  & 83.42  & 88.37  & 89.51  & 88.94  & 88.83  & 90.34  & 89.59  & 88.54  & 89.76  & 89.15  & 90.21  & 57.36  & 73.79  \\
    \hline
    1     & 5     & 89.40  & 83.89  & 86.65  & 85.47  & 92.40  & 88.93  & 81.74  & 85.74  & 83.74  & 89.27  & 89.76  & 89.51  & 88.88  & 90.27  & 89.58  & 88.95  & 89.97  & 89.46  & 85.74  & 93.57  & 89.65  \\
    \hline
    1     & 6     & 89.60  & 42.38  & 65.99  & 51.07  & 29.42  & 40.25  & 80.43  & 45.04  & 62.74  & 89.36  & 48.30  & 68.83  & 88.22  & 45.59  & 66.91  & 88.66  & 46.04  & 67.35  & 82.45  & 34.75  & 58.60  \\
    \hline
    \multicolumn{2}{|c|}{\textbf{avg}} & \multicolumn{3}{c|}{79.56 } & \multicolumn{3}{c|}{69.62 } & \multicolumn{3}{c|}{75.27 } & \multicolumn{3}{c|}{84.15 } & \multicolumn{3}{c|}{84.44 } & \multicolumn{3}{c|}{84.39 } & \multicolumn{3}{c|}{70.48 } \\
    \hline
    7     & 7     & 14.81  & 18.30  & 16.55  & 8.86  & 11.68  & 10.27  & 16.13  & 11.01  & 13.57  & 14.90  & 18.79  & 16.85  & 19.14  & 19.65  & 19.40  & 12.85  & 13.21  & 13.03  & 19.26  & 9.66  & 14.46  \\
    \hline
    7     & 8     & 12.27  & 20.72  & 16.49  & 13.68  & 40.97  & 27.33  & 14.42  & 38.07  & 26.25  & 17.57  & 49.30  & 33.44  & 18.84  & 42.04  & 30.44  & 13.56  & 52.32  & 32.94  & 9.96  & 26.54  & 18.25  \\
    \hline
    7     & 9     & 13.56  & 6.08  & 9.82  & 13.82  & 3.14  & 8.48  & 14.77  & 5.91  & 10.34  & 15.31  & 11.00  & 13.15  & 19.54  & 10.25  & 14.90  & 17.96  & 7.91  & 12.93  & 12.67  & 10.60  & 11.63  \\
    \hline
    7     & 10    & 15.37  & 14.66  & 15.01  & 13.95  & 19.75  & 16.85  & 13.55  & 27.58  & 20.56  & 15.98  & 28.87  & 22.42  & 18.55  & 27.93  & 23.24  & 12.70  & 28.44  & 20.57  & 10.51  & 34.00  & 22.25  \\
    \hline
    7     & 11    & 16.96  & 14.06  & 15.51  & 14.95  & 6.21  & 10.58  & 13.11  & 10.05  & 11.58  & 15.58  & 14.04  & 14.81  & 18.48  & 13.66  & 16.07  & 12.78  & 6.91  & 9.84  & 10.82  & 12.22  & 11.52  \\
    \hline
    7     & 12    & 14.11  & 16.91  & 15.51  & 8.09  & 26.59  & 17.34  & 14.38  & 24.16  & 19.27  & 18.60  & 29.03  & 23.82  & 19.23  & 29.93  & 24.58  & 12.92  & 27.44  & 20.18  & 7.91  & 26.55  & 17.23  \\
    \hline
    \multicolumn{2}{|c|}{\textbf{avg}} & \multicolumn{3}{c|}{14.82 } & \multicolumn{3}{c|}{15.14 } & \multicolumn{3}{c|}{16.93 } & \multicolumn{3}{c|}{20.75 } & \multicolumn{3}{c|}{21.44 } & \multicolumn{3}{c|}{18.25 } & \multicolumn{3}{c|}{15.89 } \\
    \hline
    \end{tabular}%
  \label{tbl2}
  \vspace{3pt}
\end{table*}
\end{scriptsize}

\subsubsection{Benchmarks}
We employ the benchmarks (in Table~\ref{tbl1}) used in previous studies – SABRE~\cite{c12_SABRE}, QSAM-Bench~\cite{c22_QASM}, RevLib~\cite{c23_revlib} and examples in~\cite{c14_BKA}. The tiny/small-sized programs have around five qubits and tens of CNOT gates; The large-sized ones have about ten qubits and hundreds of CNOT gates. For today’s quantum chips, using these programs can be sufficient to validate our work.

\subsubsection{Comparisons}
\noindent\textbf{Separate execution.} It maps and executes each program in a workload separately using the algorithm with the highest optimization level in qiskit~\cite{c13_qiskit}. These cases are without interference caused by multi-programming.

\noindent\textbf{Multi-programming baseline.} It uses the policy proposed in~\cite{c8_multi}, which generates initial mapping for concurrent quantum programs with FRP strategy and generates mapping transition with the enhanced noise-aware SABRE strategy. It is denoted as Baseline hereafter.

\noindent\textbf{SABRE.} Multiple programs are merged into one quantum circuit and mapped using SABRE~\cite{c12_SABRE}. SABRE is a noise-unaware approach for reducing SWAP overheads.

{\color{black} \noindent\textbf{QuCloud.} It is designed for mapping concurrent quantum programs \cite{c52_qucloud}. Crosstalk and the profiling results of the quantum programs (Sec. \ref{part:profiling}) are not considered in QuCloud.}

We show the breakdown of our approach, i.e., CDAP-only and X-SWAP-only, separately. We also show the effectiveness of our approach that enables both CDAP and X-SWAP at the same time. CDAP-only employs the same mapping transition approach with SABRE. The X-SWAP-only strategy employs the identical initial mapping strategy as SABRE.

\subsection{Evaluation results}
\subsubsection{Evaluations on Fidelity}
{
\color{black} We use tiny-sized and small-sized benchmarks for fidelity evaluation. We show PST of two-program workloads executed on IBMQ toronto in Table~\ref{tbl2}. The calibration data are the same for all experiments in Table \ref{tbl2}. The combination of two programs can double the throughput of the quantum computer. However, multi-programming may impair reliability due to resource conflicts and crosstalk errors. On average, the PST of Baseline and SABRE is lower than that in separate execution cases. Our approach incurs less fidelity reduction and can even achieve higher fidelity than separate execution cases. Generally, our approach provides a more reliable result and incurs less fidelity reduction. The average PST of this work (i.e., QuCloud+), QuCloud, separate execution, SABRE, and multi-programming baseline for tiny-sized workloads are 84.44\%, 84.15\%, 79.56\%, 69.62\%, and 75.27\%, respectively. Higher is better. For small-sized workloads, they are 21.44\%, 20.75\%, 14.82\%, 15.14\% and 16.93\%, respectively. The fidelity of QuCloud+ outperforms SABRE, the multi-programming baseline, and QuCloud by 10.56\%, 6.84\%, and 0.49\%, on average, respectively. SABRE has the lowest average fidelity as it doesn't consider noise. QuCloud+ is crosstalk aware; therefore, it improves the fidelity by 0.49\% on average compared with QuCloud.

The benefit of our approach mainly comes from CDAP. CDAP improves the fidelity by providing a better initial mapping, which makes the fidelity in multi-programming close to or even exceed that in separate execution cases. For example, as shown in Table \ref{tbl2}, in the benchmark combination of 1 and 1, i.e., bv\_n3 and bv\_n3, CDAP-only improves the fidelity significantly by 15.74\% over the multi-programming baseline by providing a better initial mapping. The underlying reasons behind the advantages of CDAP over multi-programming baseline include - (1) Gates performed on a reliable region have a lower error rate. (2) A better initial allocation reduces mapping transition SWAP cost. On average, the CDAP-only strategy reduces the adverse impact of multi-programming and outperforms the baseline by 5.22\% in fidelity.

X-SWAP-only doesn't exhibit significant advantages on fidelity improvement for the following reasons. (1) Few SWAPs are needed to map small-sized quantum programs; the number of gates saved by X-SWAP is small. Thus the advantages of X-SWAP are not reflected in fidelity. (2) The allocation of the quantum programs may not be adjacent, so inter-program SWAPs are unlikely to be enabled. X-SWAP performs better on chips with more qubits. X-SWAP can also leverage the profiling results to reduce additional SWAPs when mapping individual circuits on 3D quantum chips.
}

\begin{scriptsize}
\begin{table*}[t!]
\color{black}
  \caption{SWAP overheads comparison of 4-program workload on IBMQ50.}
  \vspace{-6pt}
  \centering
  \setlength{\tabcolsep}{2.7pt}
    \begin{tabular}{|cc|c|c|c|c|c|c|c|c|c|c|c|c|}
    \hline
    \multicolumn{2}{|c|}{\textbf{Original circuits}} & \multicolumn{2}{c|}{\textbf{SABRE}} & \multicolumn{2}{c|}{\textbf{Baseline}} & \multicolumn{2}{c|}{\textbf{QuCloud}} & \multicolumn{2}{c|}{\textbf{QuCloud+}} & \multicolumn{2}{c|}{\textbf{CDAP-only}} & \multicolumn{2}{c|}{\textbf{X-SWAP-only}} \\
    \hline
    \multicolumn{1}{|c|}{\textbf{ID}} & \textbf{CNOTs} & \textbf{depth} & \textbf{CNOTs$_{add}$} & \textbf{depth} & \textbf{CNOTs$_{add}$} & \textbf{depth} & \textbf{CNOTs$_{add}$} & \textbf{depth} & \textbf{CNOTs$_{add}$} & \textbf{depth} & \textbf{CNOTs$_{add}$} & \textbf{depth} & \textbf{CNOTs$_{add}$} \\
    \hline
    \multicolumn{1}{|c|}{13,14,15,16} & 716   & 825   & 789   & 864   & 648   & 1150  & 645   & 606   & 270   & 552   & 333   & 870   & 639 \\
    \hline
    \multicolumn{1}{|c|}{17,18,19,20} & 667   & 976   & 630   & 904   & 837   & 1026  & 678   & 704   & 468   & 633   & 378   & 825   & 678 \\
    \hline
    \multicolumn{1}{|c|}{13,14,22,28} & 572   & 789   & 714   & 484   & 423   & 1001  & 531   & 656   & 381   & 506   & 327   & 532   & 354 \\
    \hline
    \multicolumn{1}{|c|}{15,16,25,26} & 721   & 859   & 630   & 995   & 774   & 1011  & 726   & 684   & 456   & 552   & 471   & 833   & 702 \\
    \hline
    \multicolumn{1}{|c|}{17,18,23,24} & 501   & 565   & 318   & 831   & 561   & 853   & 447   & 668   & 393   & 640   & 393   & 722   & 384 \\
    \hline
    \multicolumn{1}{|c|}{19,20,21,22} & 453   & 639   & 492   & 616   & 426   & 616   & 342   & 506   & 231   & 457   & 249   & 647   & 285 \\
    \hline
    \multicolumn{1}{|c|}{17,19,22,24} & 375   & 425   & 372   & 460   & 384   & 389   & 264   & 338   & 237   & 376   & 273   & 347   & 279 \\
    \hline
    \multicolumn{1}{|c|}{18,20,21,23} & 579   & 841   & 663   & 642   & 498   & 1258  & 555   & 699   & 357   & 682   & 381   & 766   & 615 \\
    \hline
    \multicolumn{1}{|c|}{13,16,20,24} & 708   & 869   & 660   & 969   & 702   & 843   & 498   & 612   & 312   & 552   & 261   & 798   & 486 \\
    \hline
    \multicolumn{1}{|c|}{14,15,21,22} & 410   & 528   & 285   & 500   & 405   & 648   & 306   & 514   & 219   & 625   & 342   & 547   & 345 \\
    \hline
    \multicolumn{1}{|c|}{13,14,23,27} & 637   & 594   & 516   & 670   & 510   & 972   & 369   & 556   & 441   & 448   & 420   & 843   & 603 \\
    \hline
    \multicolumn{1}{|c|}{15,16,23,24} & 637   & 549   & 366   & 971   & 606   & 1090  & 519   & 627   & 237   & 552   & 294   & 929   & 621 \\
    \hline
    \multicolumn{2}{|c|}{\textbf{avg}} & 704.9  & 536.3  & 742.2  & 564.5  & 904.8  & 490.0  & 597.5  & 333.5  & 547.9  & 343.5  & 721.6  & 499.3  \\
    \hline
    \end{tabular}%
  \label{tbl3}
\end{table*}
\end{scriptsize}

\begin{scriptsize}
\begin{table*}[tp]
  \vspace{3pt}
  \caption{SWAP overheads comparison of one circuit on the 3D quantum chip.}
  \vspace{-6pt}
  \centering
  \setlength{\tabcolsep}{1.2pt}
    
    \begin{tabular}{|c|c|c|c|c|c|c|c|c|c|c|c|c|c|c|}
    \hline
    \multicolumn{4}{|c|}{\textbf{Original circuit}} & \multicolumn{2}{c|}{\textbf{SABRE}} & \multicolumn{2}{c|}{\textbf{Baseline}} & \multicolumn{2}{c|}{\textbf{QuCloud}} & \multicolumn{2}{c|}{\textbf{QuCloud+}} & \multicolumn{3}{c|}{\textbf{Comparison ($\Delta$CNOTs)}} \\
    \hline
    \textbf{ID} & \textbf{name} & \textbf{n} & \textbf{CNOTs} & \textbf{depth} & \textbf{CNOTs$_{add}$} & \textbf{depth} & \textbf{CNOTs$_{add}$} & \textbf{depth} & \textbf{CNOTs$_{add}$} & \textbf{depth} & \textbf{CNOTs$_{add}$} & \textbf{vs.  SABRE} & \textbf{vs. Baseline} & \textbf{vs. QuCloud} \\
    \hline
    13    & aj-e11\_165 & 5     & 69    & 205   & 99    & 226   & 117   & 211   & 90    & 202   & 93    & -6    & -24   & 3 \\
    \hline
    14    & alu-v2\_31 & 5     & 198   & 625   & 336   & 626   & 330   & 607   & 273   & 576   & 258   & -78   & -72   & -15 \\
    \hline
    15    & 4gt4-v0\_72 & 6     & 113   & 348   & 180   & 357   & 168   & 349   & 171   & 350   & 165   & -15   & -3    & -6 \\
    \hline
    16    & sf\_276 & 6     & 336   & 1068  & 528   & 1060  & 549   & 1044  & 474   & 1051  & 465   & -63   & -84   & -9 \\
    \hline
    17    & alu-bdd\_288 & 7     & 38    & 124   & 63    & 111   & 60    & 114   & 51    & 113   & 57    & -6    & -3    & 6 \\
    \hline
    18    & ex2\_227 & 7     & 275   & 885   & 453   & 873   & 435   & 844   & 417   & 842   & 387   & -66   & -48   & -30 \\
    \hline
    19    & ham7\_104 & 7     & 149   & 480   & 234   & 456   & 228   & 454   & 204   & 434   & 213   & -21   & -15   & 9 \\
    \hline
    20    & C17\_204 & 7     & 205   & 662   & 324   & 642   & 321   & 629   & 333   & 586   & 297   & -27   & -24   & -36 \\
    \hline
    21    & bv\_n10 & 10    & 9     & 26    & 9     & 33    & 21    & 29    & 12    & 22    & 18    & 9     & -3    & 6 \\
    \hline
    22    & ising\_model\_10 & 10    & 90    & 93    & 0     & 137   & 45    & 145   & 51    & 125   & 15    & 15    & -30   & -36 \\
    \hline
    23    & qft\_10 & 10    & 90    & 162   & 66    & 169   & 81    & 152   & 129   & 184   & 66    & 0     & -15   & -63 \\
    \hline
    24    & sys6-v0\_111 & 10    & 98    & 217   & 153   & 215   & 189   & 210   & 183   & 211   & 153   & 0     & -36   & -30 \\
    \hline
    25    & sym9\_146 & 12    & 148   & 332   & 237   & 398   & 249   & 353   & 240   & 392   & 234   & -3    & -15   & -6 \\
    \hline
    26    & rd53\_311 & 13    & 124   & 319   & 219   & 356   & 237   & 327   & 231   & 339   & 204   & -15   & -33   & -27 \\
    \hline
    27    & qft\_16 & 16    & 240   & 342   & 207   & 389   & 237   & 328   & 327   & 381   & 210   & 3     & -27   & -117 \\
    \hline
    28    & cnt3-5\_180 & 16    & 215   & 522   & 354   & 544   & 408   & 577   & 369   & 516   & 312   & -42   & -96   & -57 \\
    \hline
    \multicolumn{4}{|c|}{\textbf{avg}} & 400.6  & 216.4  & 412.0  & 229.7  & 398.3  & 222.2  & 395.3  & 196.7  & -19.7  & -33.0  & -25.5  \\
    \hline
    \end{tabular}%
  \label{tbl4}
  \vspace{3pt}
\end{table*}
\end{scriptsize}

\subsubsection{Evaluations on SWAP Overheads}
X-SWAP performs better in an enlarged SWAP search space when larger-sized programs are co-located on a quantum chip with more qubits. We evaluate SWAP overheads of 4-program workloads on IBMQ50 by comparing the number of additional CNOT gates and circuit depth. The workloads are randomly selected aiming to cover as many orthogonal program combinations as possible. Experimental results are shown in Table \ref{tbl3}. The Original circuits columns show the ID of quantum programs and the number of CNOTs in the workload. The SABRE, Baseline, QuCloud, QuCloud+, CDAP-only, and X-SWAP-only columns show the post-mapping circuit depth and the number of additional CNOTs inserted.

For SABRE~\cite{c12_SABRE}, using the reverse traversal technique and the heuristic search scheme, it tries to minimize the number of SWAPs inserted for mapping quantum programs. However, SABRE cannot achieve the optimal solution when mapping multi-programming workloads, as the locality is not exploited. Program qubits with interactions may be allocated far apart because the reverse traversal starts from a randomly generated initial mapping. By contrast, multi-programming baseline takes the locality into account when partitioning qubits, but resource conflicts are introduced between co-located quantum programs, leading to redundant SWAPs. Therefore, the experimental results show that the number of additional CNOT gates used by multi-programming baseline is 5.3\% higher SABRE, on average. 

On average, CDAP-only saves 35.9\% and 39.1\% additional CNOT gates and reduces 22.3\% and 26.2\% post-mapping circuit depth compared with SABRE and Baseline, respectively. CDAP reduces the SWAP overheads mainly for two reasons. (1) CDAP allocates tightly inter-connected qubit regions to each quantum program (Sec. \ref{sec:partition}), thereby reducing SWAP costs. (2) CDAP leverage high-degree physical qubits to reduce the SWAP overheads. X-SWAP-only employs the identical initial mapping strategy as SABRE. On average, X-SWAP-only reduces the number of additional CNOT gates by 6.9\% and 11.6\% compared with SABRE and Baseline. The reasons are multi-folds. (1) X-SWAP uses the inter-program SWAPs, taking shortcuts and saving SWAP overheads. (2) X-SWAP select gates for extended set according to the \textit{involvement list}, enhancing the look-ahead ability of the heuristic cost function.

In our design, CDAP generates a reliable and closely inter-connected initial mapping; X-SWAP helps to reduce the SWAP overheads. CDAP and X-SWAP work together to benefit the performance – reducing the additional CNOT gates by 37.8\% compared with SABRE, and 40.9\% compared with Baseline. The circuit depth is reduced by 15.2\% and 19.5\% compared with SABRE and Baseline, respectively. By exploiting the profiling results, QuCloud+ reduces 31.9\% additional CNOTs and 34.0\% post-mapping circuit depth compared to QuCloud. More results can be found in Table \ref{tbl3}.

Moreover, our work exhibits scalability. It reduces the SWAP overheads for 4-program workloads on IBMQ50. It also can be used on a larger quantum chip with more qubits, because – (1) The community detection approach in CDAP has been proved to be effective for large networks. (2) X-SWAP reduces SWAP overheads when quantum programs are mapped adjacently. They both work well regardless of the scale of a specific quantum chip.

\subsubsection{Evaluations on the 3D Quantum Chip }
{\color{black}
We map each large-sized circuit in Table \ref{tbl1} onto the 3D quantum chip with the topology architecture illustrated in Figure \ref{fig:3D}. Table \ref{tbl4} shows the results of circuit depth after mapping and number of additional gates inserted. The $\Delta$CNOTs columns show how many additional gates QuCloud+ saves compared to the three other competing mechanisms. Negative values indicate that QuCloud+ inserts fewer additional gates during the mapping transition.

QuCloud+ has the least additional gates. It saves 9.1\%, 14.4\% and 11.5\% CNOT gates compared with SABRE, Baseline and QuCloud, on average, respectively. QuCloud+ win advantages on the circuit depth. The benefits come from the qubit allocation and mapping transition mechanisms based on profiling results of the quantum circuit. During initial mapping generation, high-degree physical qubits on the 3D quantum chip are fully utilized, allowing more program qubits to have direct interconnections. During mapping transition, only the latest CNOT gates that consecutively involve each program qubit are added to the extended set. Compared to blindly selecting a fixed number of CNOTs in topological order in prior work, this approach enhances the look-ahead ability of the extended set, thus reducing additional gates inserted.
}

\subsubsection{Evaluations on the QuCloud+ Scheduler}

{\color{black}
We build three task queues to evaluate the QuCloud+ scheduler. Each queue contains 100 quantum programs randomly selected from Table~\ref{tbl1}. We use the QuCloud+ scheduler to schedule the workloads with the estimated fidelity (EPST) violation threshold $\varepsilon$ ranging from 0.05 to 0.20. Then the workloads are executed on the IBMQ toronto QASM simulator. The Probability of a Successful Trial (PST) and Trial Reduction Factor (TRF) averaged across the 3 task queues are shown in Figure~\ref{fig:14}. Figure~\ref{fig:14} also shows the performance in separate execution cases and the randomly selected two-programmed combination cases.

\begin{figure}[tbp]
\centering
\includegraphics[width=0.9\linewidth]{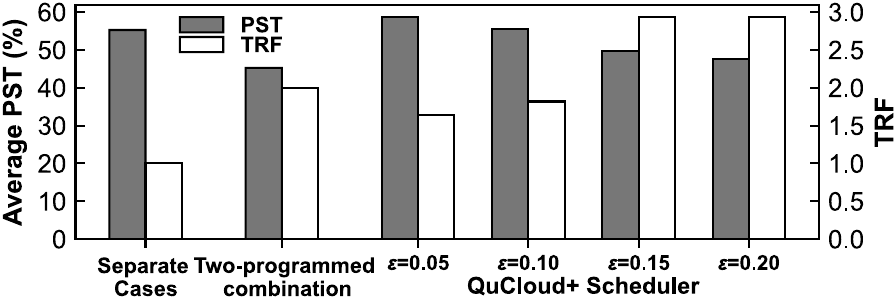}
\vspace{-0.2cm}
\caption{Performance of the QuCloud+ scheduler. PST and TRF stand for fidelity and throughput, respectively. Higher is better for both. The increase of $\varepsilon$ leads to higher throughput, but may cause fidelity reduction.}
\label{fig:14}
\end{figure}

Separate execution can support the best average PST of 55.2\% with a TRF of 1 (no parallelism). Randomly selected two-programmed combination cases have the average PST of 45.2\%, the TRF is 2 (i.e., 2 programs in parallel all the time). By contrast, QuCloud+ scheduler can reach the highest TRF of 2.94 when $\varepsilon$ is 0.15 (i.e., only multi-programming cases leading to less than 15\% estimated fidelity reduction could be scheduled). In this case, the average fidelity of the workloads is 49.8\%, which is only 5.4\% worse than the separate cases, but still 4.6\% higher than the randomly selected two-programmed combination cases. The TRF is 2.94, indicating the throughput is improved by 194\% compared to separate execution cases. The experimental results show that QuCloud+ scheduler can provide better solutions to balance the quantum computers' throughput and fidelity.
}
\section{Conclusion}

Quantum computers attract more and more attention. With the trend of putting everything on the cloud, quantum computers face the problems of resource under-utilization, lower fidelity, higher error rates, etc. Our work presents a new qubit mapping policy for multi-programming cases, improving the fidelity and resource utilization when multiple quantum programs are running on a specific quantum chip. Our approach outperforms the state-of-the-art multi-programming strategy by improving the fidelity and reducing the SWAP overheads. As multi-programming is gaining importance in the cloud, we hope our efforts could help future researchers in the related field.



\ifCLASSOPTIONcaptionsoff
  \newpage
\fi



\begin{IEEEbiographynophoto}{Lei Liu} is a professor in Beihang university and ICT, CAS, where he leads the Sys-Inventor Lab. His research interests include quantum computing, OS, memory architecture and computer architecture. His efforts are published in ISCA, HPCA, PACT, IEEE TC, TPDS, ACM TACO, ICCD, and etc.
\end{IEEEbiographynophoto}

\begin{IEEEbiographynophoto}{Xinglei Dou} is a student member of Sys-Inventor Lab in ICT, CAS. His work includes the quantum computing and OS design. He is supervised by Lei Liu.
\end{IEEEbiographynophoto}

\end{document}